\documentclass[iop]{emulateapj}
\usepackage{graphics}
\usepackage{amsmath}

\newcommand{\Pstar}{P{\text{*}}}

\begin{document}

\title{Turbulence, transport and waves in Ohmic dead zones}

\author{Daniel Gole\altaffilmark{1,2}, Jacob B. Simon\altaffilmark{1,3,4}, Stephen H. Lubow\altaffilmark{5} and Philip J. Armitage\altaffilmark{1,2}}
 
\begin{abstract}
We use local numerical simulations to study a vertically stratified accretion disk with a resistive mid-plane that damps magnetohydrodynamic (MHD) turbulence.  This is an idealized model for the dead zones that may be present at some radii in protoplanetary and dwarf novae disks. We vary the relative thickness of the dead and active zones to quantify how forced fluid motions in the dead zone change. We find that the residual Reynolds stress near the mid-plane decreases with increasing dead zone thickness, becoming negligible in cases where the active to dead mass ratio is less than a few percent. This implies that purely Ohmic dead zones would be vulnerable to episodic accretion outbursts via the mechanism of \cite{martin11}. We show that even thick dead zones support a large amount of kinetic energy, but this energy is largely in fluid motions that are inefficient at angular momentum transport. Confirming results from \cite{oishi09}, the perturbed velocity field in the dead zone is dominated by an oscillatory, vertically extended circulation pattern with a low frequency compared to the orbital frequency. This disturbance has the properties predicted for the lowest order r mode in a hydrodynamic disk. We suggest that in a global disk similar excitations would lead to propagating waves, whose properties would vary with the thickness of the dead zone and the nature of the perturbations (isothermal or adiabatic).  Flows with similar amplitudes would buckle settled particle layers and could reduce the efficiency of pebble accretion.
\end{abstract} 


\altaffiltext{1}{JILA, University of Colorado and NIST, 440 UCB, Boulder, CO 80309, USA}
\altaffiltext{2}{Department of Astrophysical and Planetary Sciences, University of Colorado, Boulder, CO 80309, USA}
\altaffiltext{3}{Department of Space Studies, Southwest Research Institute, Boulder, CO 80302, USA}
\altaffiltext{4}{Sagan Fellow}
\altaffiltext{5}{Space Telescope Science Institute, 3700 San Martin Drive, Baltimore, MD 21218, USA}

\section{Introduction} 
Turbulence and angular momentum transport in well-ionized accretion disks are generated by the 
magnetorotational instability \citep[MRI;][]{balbus91,balbus98}. In weakly ionized disks, however, 
the efficiency of the MRI is modified by non-ideal magnetohydrodynamic effects --- Ohmic diffusion, 
ambipolar diffusion, and the Hall effect \citep{blaes94}. These processes are predicted to impact the dynamics 
of accretion in the cool outskirts of cataclysmic variable \citep{gammie98} and Active Galactic 
Nuclei disks \citep{menou01}, and are dominant across most radii in protoplanetary disks 
\citep[for reviews, see e.g.][]{armitage11,turner14}.

The simplest explicit model for the effect of the non-ideal MHD terms was introduced by \citet{gammie96}, who considered the damping of the MRI by Ohmic diffusion in a disk with scale height $h$, Alfv\'en speed $v_A$ and resistivity $\eta$. By equating the minimum Ohmic damping rate, $\eta / h^2$, with the linear MRI growth rate on the same scale, $v_A/h$, he argued that local turbulence would be damped for resistivities exceeding $\eta_{\rm crit} \sim h v_A$. For AU-scales in 
protoplanetary disks this resistivity corresponds to an ionization fraction of the order of $10^{-13}$, which is larger 
than that expected at the disk mid-plane via thermal or plausible non-thermal ionization sources.  These
considerations led to the prediction of an Ohmic dead zone around the mid-plane in which $\eta > \eta_{\rm crit}$.  In this model, accretion preferentially occurs in non-thermally ionized surface layers in which the MRI drives turbulence. Simulations by \citet{fleming03}, using a similarly simplified model that included only Ohmic diffusion, verified Gammie's qualitative disk structure. However, they also showed that waves generated within the MRI-active zones could leak into the dead zone, producing turbulence and purely hydrodynamic stresses.

The presence of a dead zone can qualitatively change the nature of protostellar accretion and planet 
formation. For accretion, the dependence of the vertically integrated stress $\alpha$ \citep{shakura73} on the surface 
density $\Sigma$ determines whether the disk, at some radius $r$, is able to attain a steady-state for a 
specified accretion rate $\dot{M}$. Simple models of Ohmic dead zones do {\em not} in general admit 
steady-states for all $\dot{M}$ \citep{gammie96}, and are potentially vulnerable to a gravo-magneto limit cycle instability 
\citep{martin11} that has been conjectured to be the origin of episodic accretion in Young Stellar 
Objects \citep{armitage01,zhu09}. For planet formation, the effective diffusivity $D$ generated by 
turbulent velocity fluctuations within the dead zone sets the thickness, and hence the density, of 
settled particle layers. Surface density fluctuations, on the other hand, generate stochastic gravitational 
torques that may dominate the excitation (and limit the growth) of planetesimals \citep{ida08,gressel11,gressel12,okuzumi13}.

Although the relative importance of the non-ideal MHD terms varies with density, temperature, and degree 
of magnetization \citep{balbus01,kunz04,xu15}, in protoplanetary disks it is often the case that multiple 
non-ideal effects are simultaneously important. At large disk radii ($\sim 30-100 \ {\rm AU}$) ambipolar 
diffusion is dominant \cite[e.g.,][]{desch04,kunz04,simon13a,simon13b}, but on AU-scales both Ohmic diffusion and the Hall 
effect modify the mid-plane dynamics \citep{desch04,kunz04,wardle07,lesur14,bai14,simon15}, while ambipolar diffusion is important in the atmosphere \citep{desch04,kunz04,wardle07,bai13}. Accurately capturing the full non-ideal physics is computationally difficult, and the numerical expense limits the feasible resolution and / or run length. Here, following \cite{fleming03} 
and \cite{okuzumi11}, we instead study the model problem of a purely Ohmic 
dead zone, which we further simplify by imposing a sharp vertical transition between the active and dead layers. 
By examining the interaction of dead zone and active zone gas in
this controlled manner, we are able to better understand the physical effects that are observed in the simulations. 
We are particularly interested in determining the functional form of $\alpha (\Sigma)$, which is the critical 
quantity for disk stability, and in studying the nature of dead zone fluid motions and their implications for 
settled particle layers.

The plan of the paper is as follows. In \S2 we describe the numerical methods and initial conditions used. \S3 presents the results of the simulations, including the scaling of dead zone properties with dead zone thickness and the nature of turbulence and large scale fluid flows in the simulated disks. \S4 discusses the implications for particle settling and growth if large-scale circulatory flows, seen in our simulations, are present in real disks. We wrap up with our conclusions in \S5.

\section{Methods}
\subsection{Numerical Algorithm}

We compute the structure of Ohmic dead zones using {\sc Athena}. {\sc Athena} is a second-order accurate Godunov flux-conservative code that uses constrained transport \cite[CT;][]{evans88} to enforce the $\nabla \cdot \mathbf{B} = 0$ constraint, the
third-order in space piecewise parabolic method (PPM) of \cite{colella84} for spatial reconstruction, and the HLLD Riemann solver \citep{miyoshi05,mignone07} to calculate numerical fluxes.  A full description of the {\sc Athena} algorithm
along with results showing the code's performance on various test problems can be found in \cite{gardiner05}, \cite{gardiner08}, and \cite{stone08}.

We employ the shearing box approximation to simulate a small, co-rotating patch of an accretion disk.  Taking the size of the shearing box to be small relative to the distance from the central object, we define Cartesian coordinates $(x, y, z)$ in terms of the cylindrical coordinates ($R,\phi,z^\prime$) such that $x=R-R_0$, $y=R_0\phi$, and $z = z^\prime$.  The box co-rotates about the central object with angular velocity $\Omega$, corresponding to the Keplerian angular velocity at $R_0$.  

The equations of resistive MHD in the shearing box approximation are,

\begin{subequations}
\begin{equation}
\frac{\partial \rho}{\partial t} + \nabla \cdot (\rho \mathbf{v}) = 0, \\
\end{equation}
\begin{equation}
\frac{\partial (\rho \mathbf{v})}{\partial t} + \nabla \cdot (\rho \mathbf{v} \mathbf{v} - \mathbf{B} \mathbf{B} + \mathbf{\bar{\bar{I}}} \Pstar) =  \\ 2 q \rho \Omega^2 x \mathbf{\hat{i}} - \rho \Omega^2 z \mathbf{\hat{k}}  - 2 \Omega \mathbf{\hat{k}} \times \rho \mathbf{v}, \\
\end{equation}
\begin{equation}
\frac{\partial \mathbf{B}}{\partial t} - \nabla \times (\mathbf{v} \times \mathbf{B}) - \eta \nabla^2 \mathbf{B} \ = 0. 
\end{equation}
\end{subequations}

\noindent Here $\rho$ is the gas density, $\mathbf{B}$ is the magnetic field, $\Pstar$ is the total pressure (related to the magnetic field and the gas pressure, $P$, via $\Pstar = P + {\mathbf{B} \cdot \mathbf{B}}/{2}$), $\mathbf{\bar{\bar{I}}}$ is the identity tensor, $\mathbf{v}$ is the velocity, $\eta$ is the Ohmic resistivity, and $q$ is the shear parameter defined by $q=-d\ln{\Omega}/d\ln{R}$, which is $3/2$ in this case of a Keplerian disk.  An isothermal equation of state with a constant sound speed $c_s$ is used,   
\begin{equation}
P = \rho c_s^2.    
\end{equation}
We use a standard set of boundary conditions appropriate for the shearing box approximation. At the azimuthal boundaries we exploit the symmetry of the disk and apply periodic boundary conditions. At the radial boundaries we  apply shearing periodic boundaries, such that quantities that are remapped from one radial boundary to the other are  shifted in the azimuthal direction by the distance the boundaries have sheared relative to each other at the given time \citep{hawley95}. In the vertical direction, we employ the modified outflow boundary condition of \cite{simon13}, which are the standard outflow boundaries but with the gas density extrapolated exponentially into the ghost zones; this method reduces the artificial build-up of toroidal magnetic flux near the vertical boundaries.  Such outflow boundaries can lead to significant mass loss, particularly in the case of strong vertical magnetic flux \citep{simon13}.  However, such drastic mass loss is not observed in our simulations, consistent with expectations from our magnetic field geometry as described below. In general, these boundary conditions can break conservation, so methods described in \cite{stone10} are used in order to keep the correct variables conserved.   Furthermore, we employ Crank-Nicholson differencing to conserve epicyclic energy to machine precision. A full discussion of these methods and {\sc Athena}'s shearing box algorithm can be found in \cite{stone10}.\\              

\subsection{Initial Conditions}

We run all simulations using a domain size of $L_x \times L_y \times L_z = 2H \times 4H \times 8H$, where $H$ is the scale height, defined here as,

\begin{equation}
H = \frac{\sqrt{2} c_s}{\Omega}.     
\label{eq_H_definition}    
\end{equation}
We assume that the disk is initially in hydrostatic equilibrium.  Balancing the vertical component of gravity from the star with the gas pressure gives a Gaussian density profile:

\begin{equation}
\rho(z) = \rho_0 \exp\left(\frac{-z^2}{H^2}\right).                  
\end{equation}

\noindent We set $\rho_0 = 1$ as the initial density at the mid-plane.  A density floor of $10^{-5}$ is applied to keep the Alfv\'en speed from getting too high (and thus the time step too small) in the region far from the mid-plane.  The density floor also prevents $\beta$ (defined below) from getting too low, which can cause numerical problems. We set 
the constant isothermal sound speed $c_s=1/\sqrt{2}$, and choose the angular velocity to be $\Omega = 1$, which gives $H = 1$.

We initialize the magnetic field to be purely toroidal with constant $\beta$ parameter of $\beta=400$, where $\beta$ is the ratio of the gas pressure to the magnetic pressure, 

\begin{equation}
\beta = \frac{2P}{B^2}.                   
\end{equation}

\noindent 
Note that a factor of $4\pi$ has been subsumed into the definition of the magnetic pressure, following the standard definition of units in {\sc Athena} \citep{stone08}.
 
The saturation level of MRI-driven turbulence is known to be a function of the net vertical magnetic flux 
\citep{hawley95,salvesen16}, which can additionally give rise to outflows \citep{bai13b} and to qualitatively 
distinct Hall physics \citep{lesur14,bai14,simon15}. Here, our 
primary goal is to study the relationship between active layer turbulence
and motions induced by this active layer in an idealized model of the dead zone, for which the actual 
saturation level of the MRI is of secondary importance. For simplicity, and to reduce the parameter space 
that needs to be studied, we therefore choose a field geometry with no net vertical magnetic flux. Consistent 
with this choice, we do not observe strong MHD outflows from our simulations.
 
In order to seed the MRI, we apply random perturbations to the gas pressure and velocity. Following the procedure used by \cite{hawley95}, the perturbations are uniformly distributed throughout the box and have zero mean.  Pressure perturbations are a maximum of $2.5 \%$ of the local pressure, and velocity perturbations are a maximum of $5\times 10^{-3}c_s$. We employ a uniform numerical resolution of 60 grid cells per $H$, giving a total resolution of $120\times240\times480$ zones per simulation.  

\subsubsection{Resistivity Profile}

Five simulations were done, varying the ratio of the surface density of the active zone $\Sigma_{\text{active}}$ to the total surface density $\Sigma_{\text{total}}$.  The surface density of the active layer is physically set by a balance between recombination and ionization from sources such as cosmic rays \citep{gammie96}, X-rays \citep{igea99}, and FUV photons \citep{perezbecker11}. In most circumstances this balance 
is achieved on time scales that are shorter than the turbulent mixing time between the surface and the interior of the disk  \citep{bai11}. It is then a reasonable approximation to ignore time-dependent ionization effects, and take the column of ionized gas to be fixed in space and time. Varying the active to total surface density ratio is thus equivalent to changing the size of the dead zone. 

Physically, the resistivity varies with column density as a smooth function whose form is set by the details 
of the ionization-recombination balance. To more accurately study the scaling of dead zone properties with dead zone size, however, we ignore this complexity and instead create a simple dead zone by imposing a sharp vertical transition in the Ohmic resistivity using an error function.  Specifically, we adopt,

\begin{equation}
\eta = {\rm MIN}(\eta_1,\eta_2)
\end{equation}

\noindent
where

\begin{equation}
\eta_1 = \frac{\eta_{\text{mid}}}{2}\left[\text{ERF}\left(\frac{z+z_{\text{crit}}}{h_\eta}\right)+1\right],                     
\end{equation}

\noindent
and

\begin{equation}
\eta_2 = \frac{\eta_{\text{mid}}}{2}\left[1-\text{ERF}\left(\frac{z-z_{\text{crit}}}{h_\eta}\right)\right].                    
\end{equation}

\noindent Here $\eta_{\rm mid}=0.005$ is the resistivity of the mid-plane, $h_{\eta}=0.2H$ controls how quickly the resistivity transitions from the dead zone at the mid-plane to the active region, and $z_{\rm crit}$ controls the height at which this transition happens.  Note that $z_{\rm crit}$ is specified by the initial conditions and remains fixed throughout the simulation, it is 
{\em not} adjusted on-the-fly to account for changes in the instantaneous vertical density distribution.

\begin{figure}[h]
\centering
\includegraphics[width=\columnwidth]{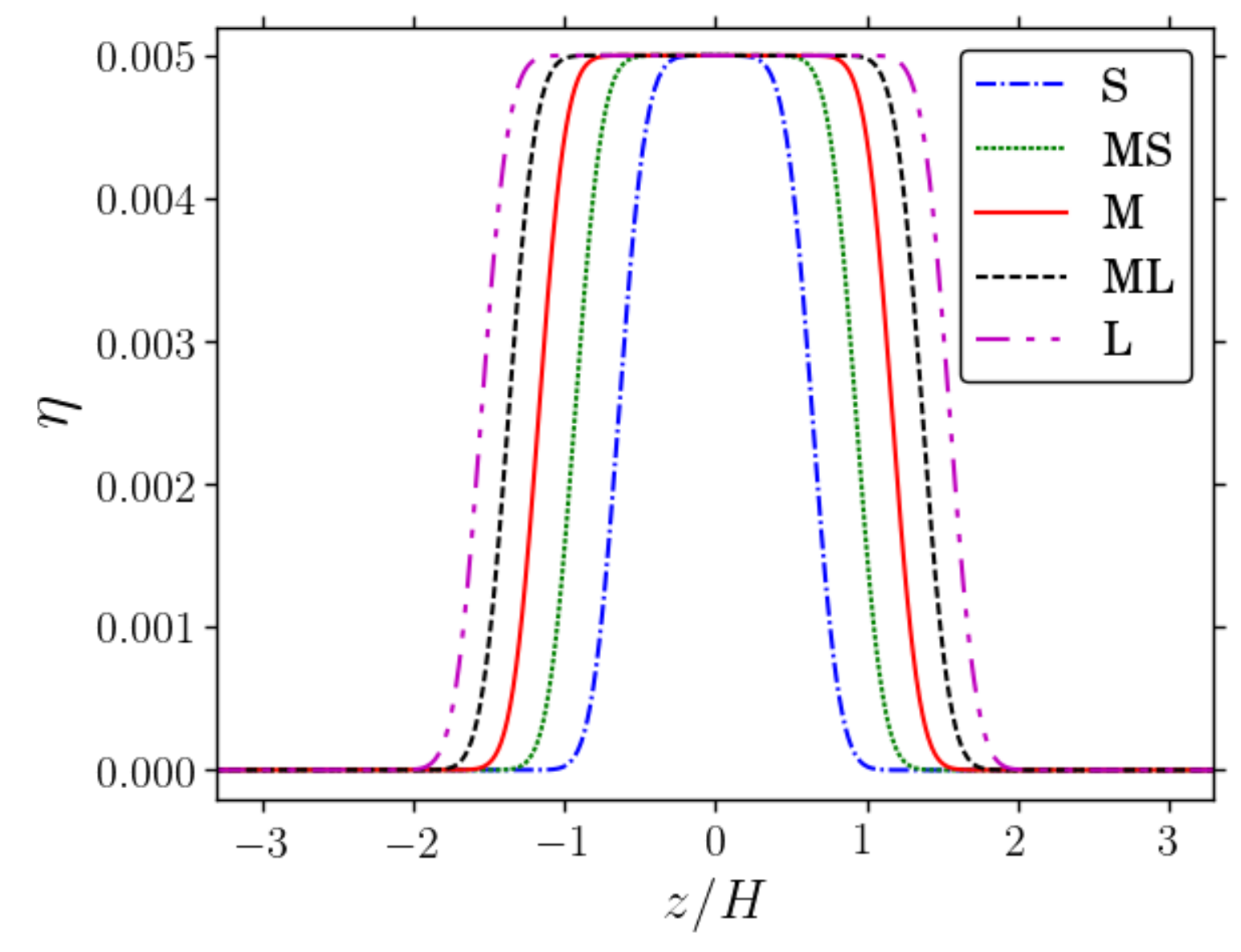}
\caption{Resistivity $\eta$ as a function of $z$ for each of our simulations, as labeled in the legend. The mass in the 
active zone as a fraction of the total mass varies between about  0.58 (for simulation ``S") and 0.029 (for simulation ``L").}
\label{figResis}
\end{figure}

Depending upon the ionization source (cosmic rays or X-rays) the ratio of active to total column at 1~AU 
in a purely Ohmic dead zone model is expected to be of the order of 0.1 to $10^{-2}$, if the underlying 
disk has a surface density comparable to the Minimum Mass Solar Nebula. Substantially smaller values 
of the order of $10^{-3}$ are possible if mass accumulates in the dead zone, as occurs in some models for 
episodic accretion \citep{armitage01,zhu09,martin11}. As we show later, however, modeling extremely thin active 
layers is numerically challenging, because the time scale for kinetic energy in the dead zone to saturate becomes 
prohibitively long. With this constraint in mind, we are able to reliably model dead zones whose active to total 
column varies between $\sim 0.5$ and $\sim 0.03$. The specific values of this ratio, along with the 
corresponding values of $z_{\rm crit}$, are shown in Table~\ref{tabSizes}.  The corresponding resistivity profiles are plotted in Figure~\ref{figResis}.

\begin{deluxetable}{l|cc}
\tablewidth{0pc}
\tablecaption{Dead Zone Size\label{tabSizes}}
\tablehead{
\colhead{Run}&
\colhead{$z_{\rm crit}/H$}&
\colhead{Active-to-Dead Mass Ratio} }
\startdata
S &  0.64 & 0.58 \\
MS &  0.93 & 0.23\\
M &  1.16 & 0.11\\
ML &  1.37 & 0.054 \\
L &  1.55 & 0.029 \\
\enddata
\end{deluxetable}

The degree to which the resistivity suppresses the MRI can be quantified by the Elsasser number, which compares the MRI growth rate to the resistive damping rate,

\begin{equation}
\Lambda = \frac{v_{az}^2}{\eta \Omega}.                     
\end{equation}

\begin{figure}[h]
\centering
\includegraphics[width=\columnwidth]{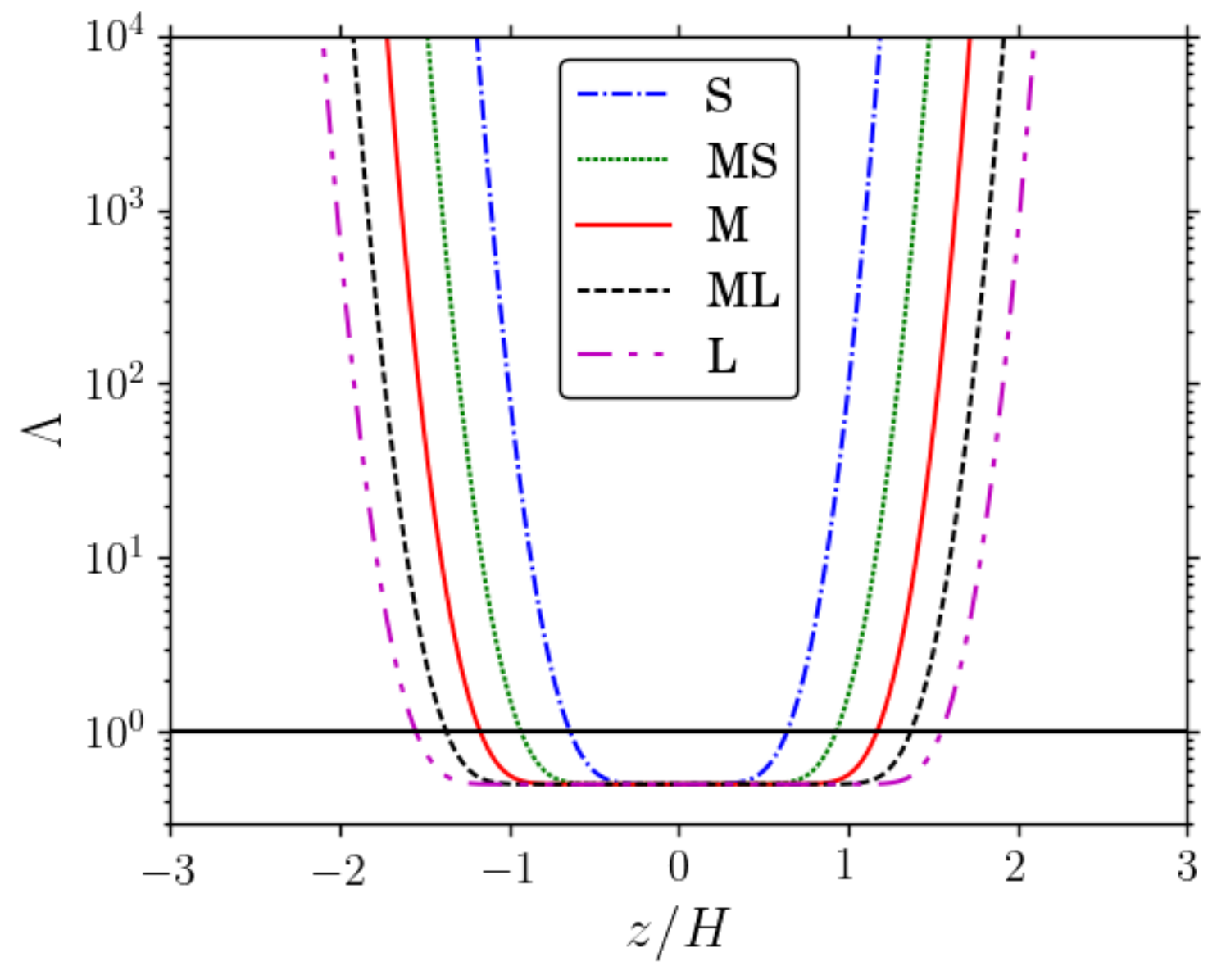}
\caption{The dependence of the Elsasser number, $\Lambda$ on $z$ for each of our simulations, as labeled via the legend.  The solid black horizontal line shows where $\Lambda=1$. The dead zone
is defined here as the region where $\Lambda < 1$.}
\label{figElsasser}
\end{figure}

\noindent We show $\Lambda$ as a function of $z$ in Fig.~\ref{figElsasser}.  Where $\Lambda>1$, the most unstable mode of the MRI is unaffected by ohmic diffusion.  Where $\Lambda<1$, the growth rate of the MRI goes as $\Lambda \Omega$ \citep{sano99}.  The initial resistivity and magnetic field strength were chosen such that $\Lambda<1$ at the mid-plane to keep the MRI suppressed in the dead zone, while maintaining a sufficiently small resistivity such that the diffusive time scale does not limit the time step. 

The local wavelength of the most unstable MRI mode is the greater of the ideal MRI wavelength and the resistive MRI wavelength, given as \citep{sano99},

\begin{align}
\lambda_{\text{ideal}} &= \frac{2\pi v_{az}}{\Omega}   \\             
\lambda_{\text{resistive}} &= \frac{\lambda_{\text{ideal}}}{\Lambda}. 
\end{align}
With a resolution of 60 zones per $H$, these MRI modes are well resolved in the initial conditions, with 20 zones per MRI wavelength at minimum in the active region and about 40 zones per wavelength in the dead zone. 

\section{Results}
\subsection{General, Large-scale Properties}

\begin{figure}[h]
\centering
\includegraphics[width=\columnwidth]{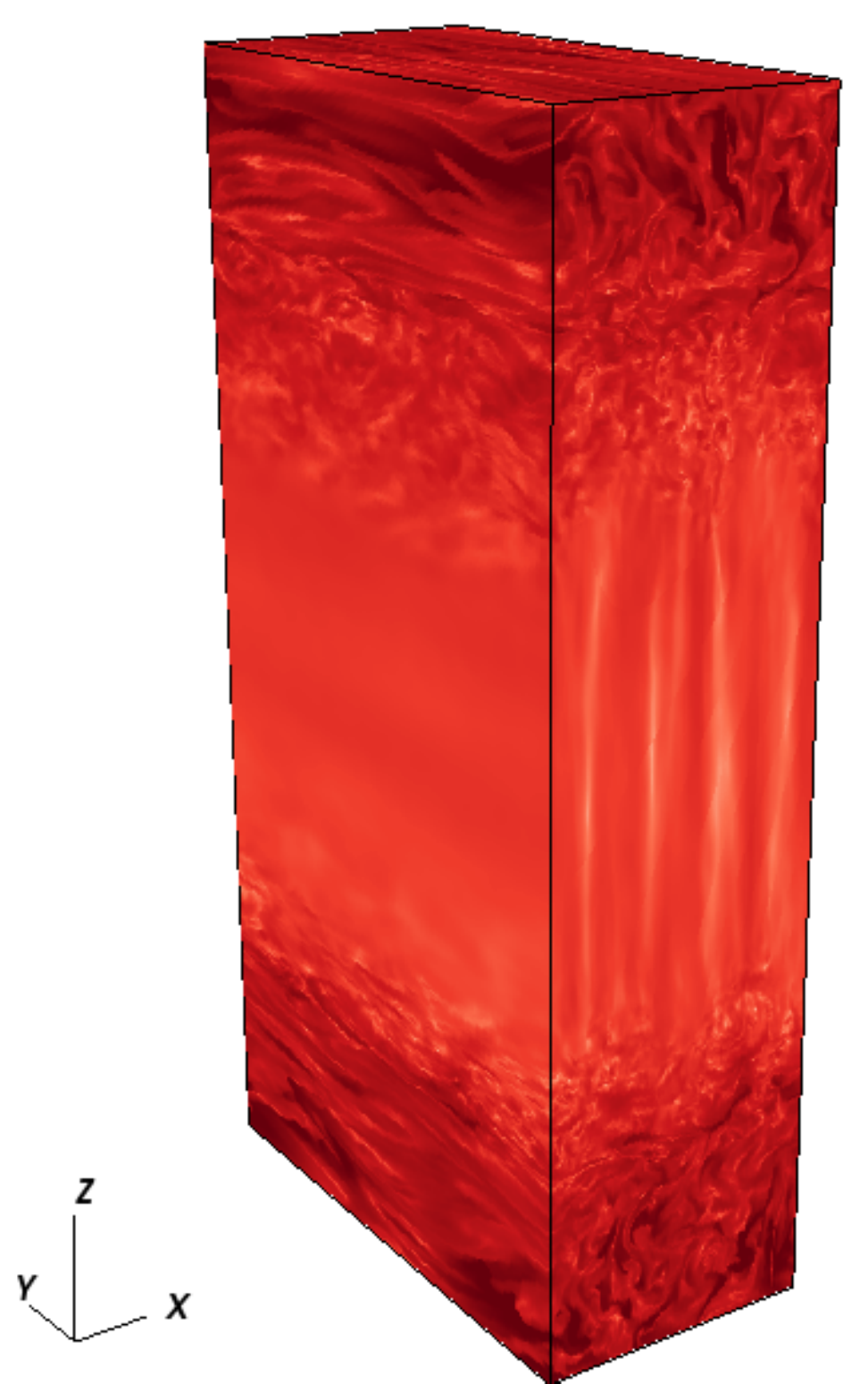}
\caption{The log of the velocity magnitude for the medium sized dead zone run, plotted with darker colors indicating higher velocity.  The outer regions are clearly turbulent, transitioning into more laminar flow at the mid-plane.  }
\label{fig3dBox}
\end{figure}

\begin{figure}[h]
\centering
\includegraphics[width=\columnwidth]{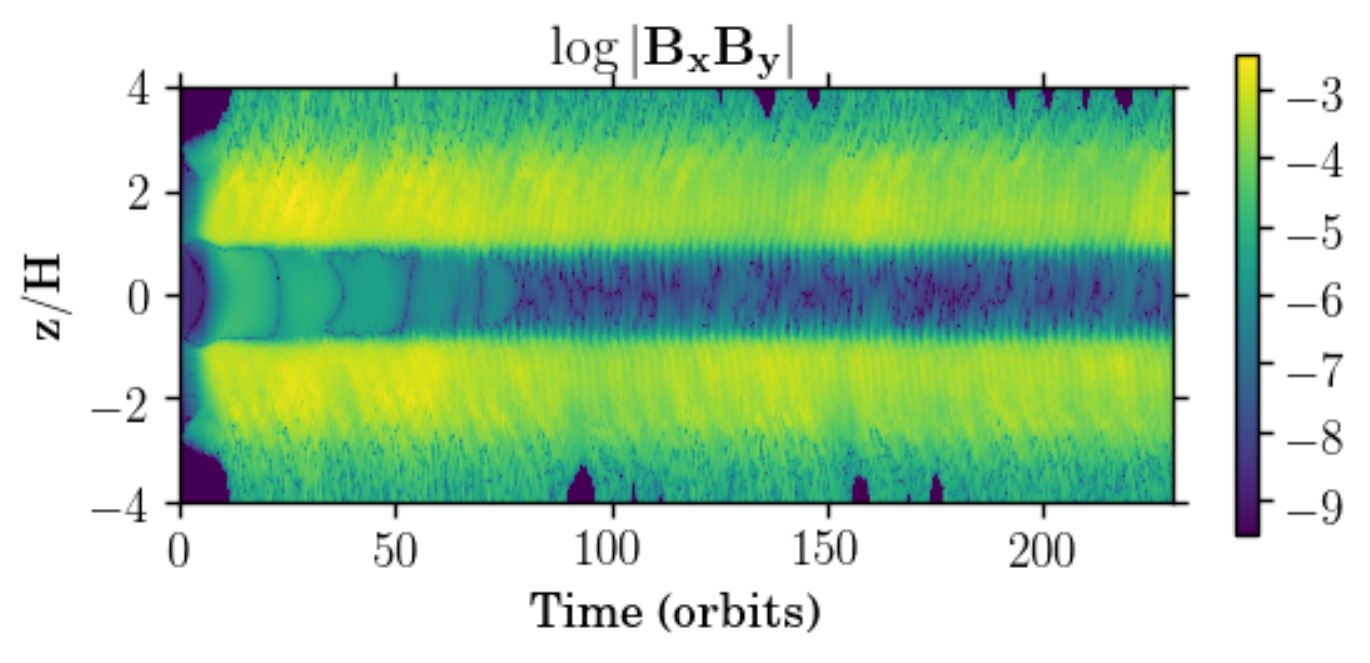}
\includegraphics[width=\columnwidth]{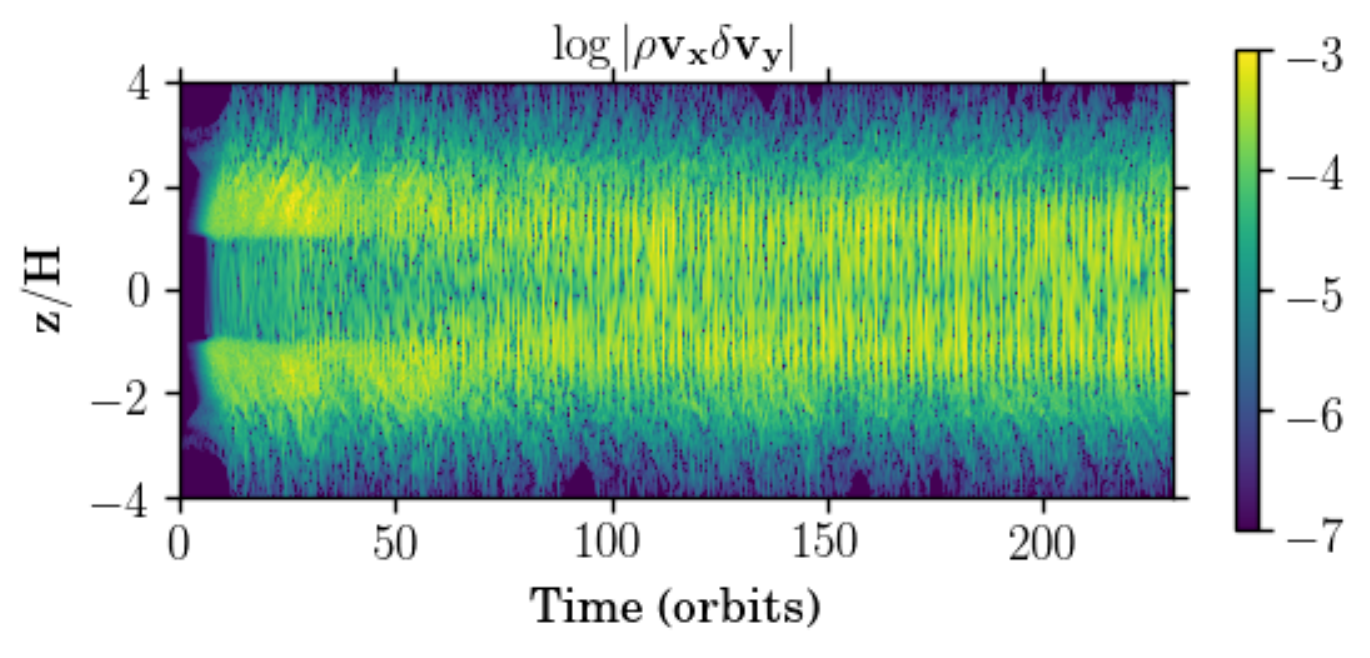}
\includegraphics[width=\columnwidth]{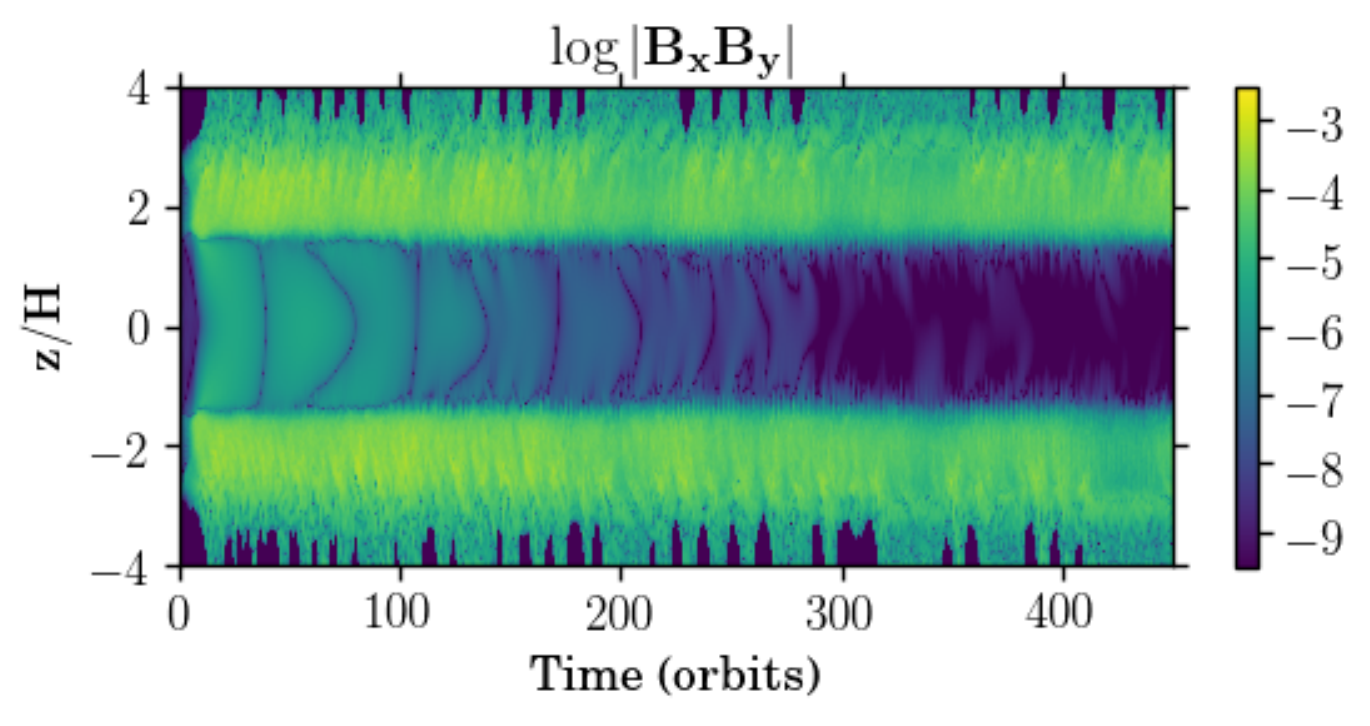}
\includegraphics[width=\columnwidth]{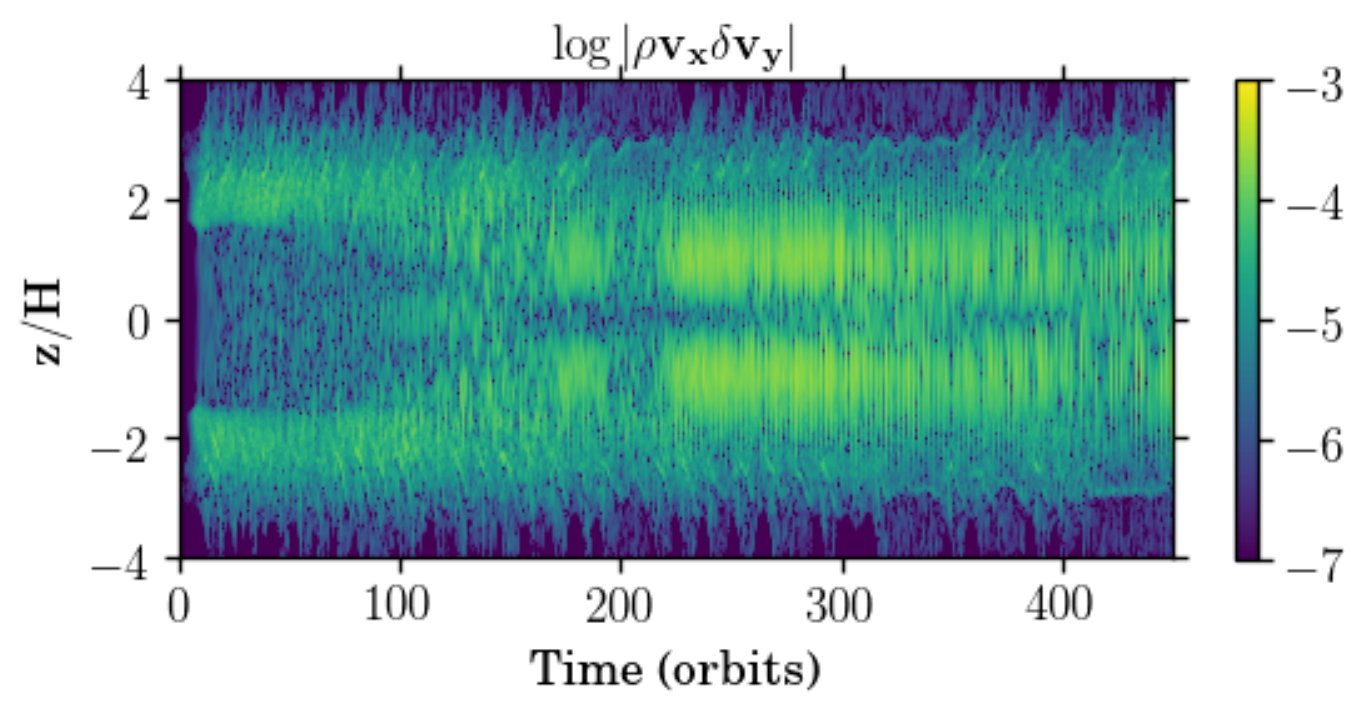}
\caption{Space-time plots of the Maxwell stress $=-B_x B_y$ (first and third plots) and Reynolds stress $=\rho v_x \delta v_y$ (2nd and 4th plots) for the small (top 2) and medium (bottom 2) dead zone sizes.  The quantities are averaged in the $x$ and $y$ directions.  Note that the color scales are kept constant for like-quantities across the runs. }
\label{figSTstress}
\end{figure}

Illustrative properties of the simulations are shown in Figure~\ref{fig3dBox} and Figure~\ref{figSTstress}. 
Figure~\ref{fig3dBox} shows a rendering of the velocity fluctuations in the simulation domain, while 
Figure~\ref{figSTstress}  shows space-time diagrams of the Maxwell and Reynolds stresses for the 
small and medium dead zone cases (note that the stress is {\em not} normalized to the mean local 
density, and hence is not equivalent to a local $\alpha$ value). 
It is clear from either figure that the imposed resistivity profile 
delivers the expected morphology of an idealized Ohmic dead zone. The Maxwell stress saturates 
rapidly to a dominant level in the active zone, and is damped to very low levels in the dead zone. 
The Reynolds stress starts low in the dead zone and then builds up over time. Fluctuations in the Reynolds stress in the dead zone ultimately build up to be as high or higher than in the active region.     

\subsection{Saturation Time-scales}

\begin{figure}[h]
\centering
\includegraphics[width=\columnwidth]{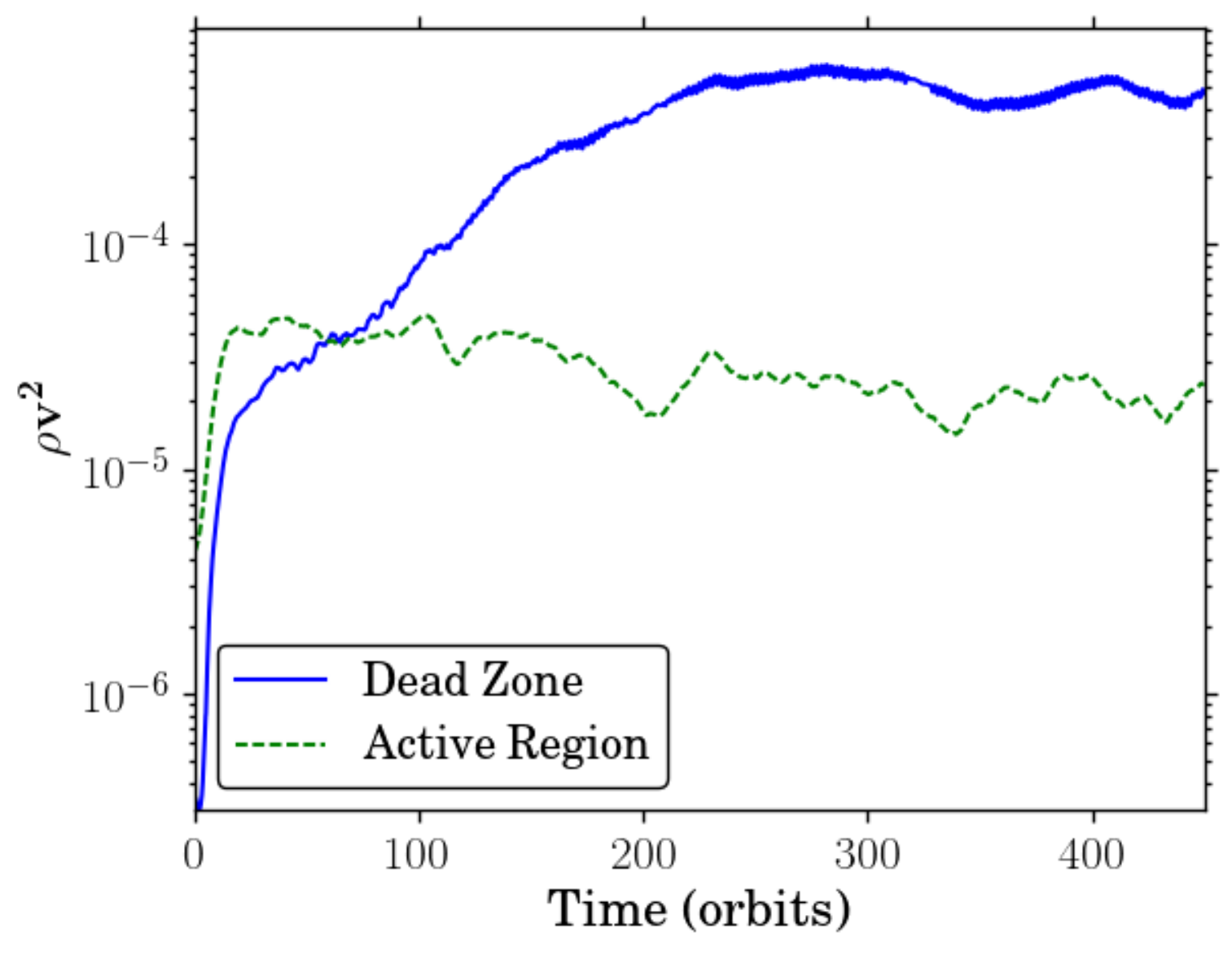}
\includegraphics[width=\columnwidth]{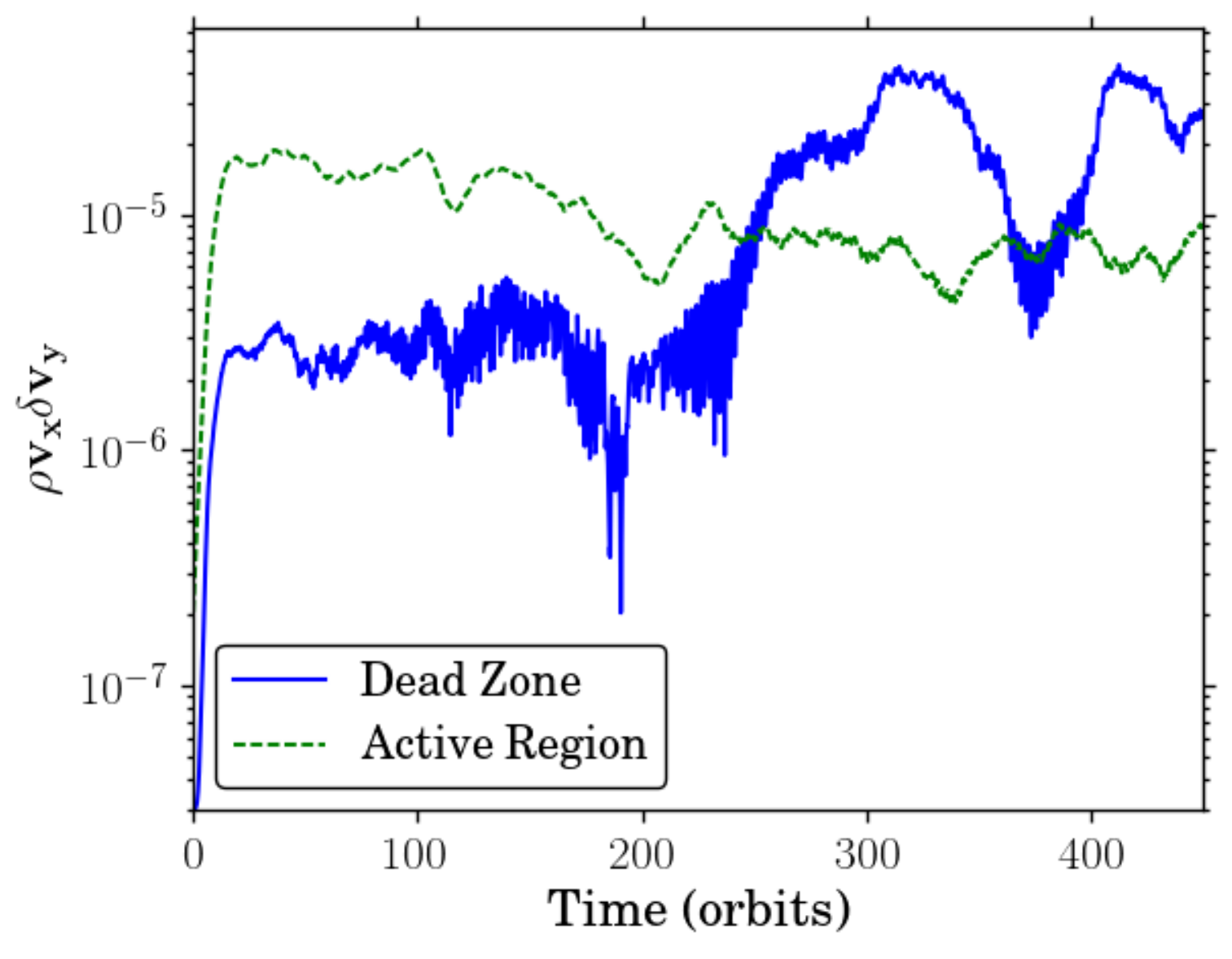}
\includegraphics[width=\columnwidth]{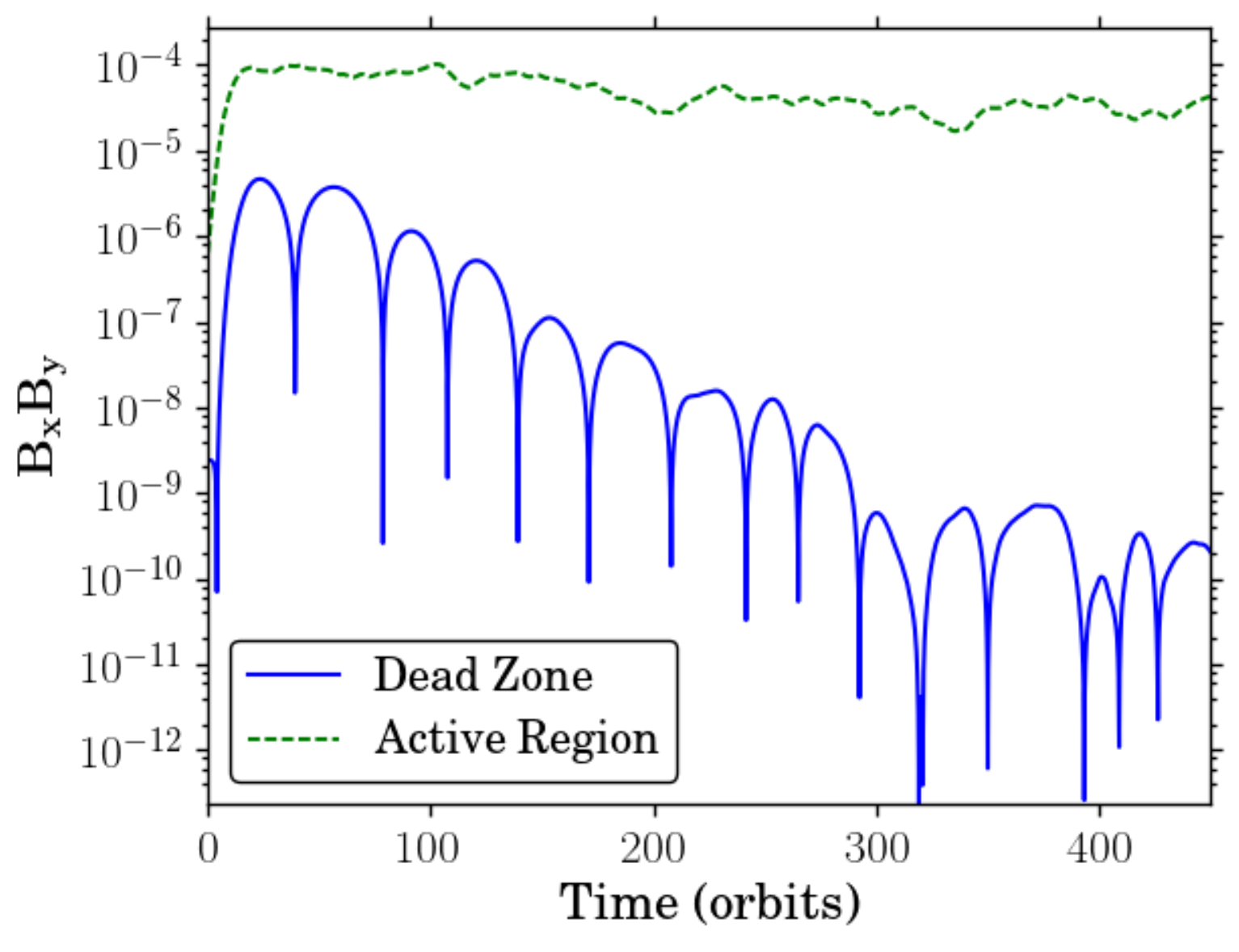}

\caption{Time evolution of the volume averaged Kinetic Energy, Reynolds Stress, and Maxwell Stress (from top to bottom), for the medium sized dead zone.  The average over the dead zone is taken from $|z/H|=0$ to $0.5$, and for the active zone from $|z|=2.0$ to $3.3$.  In addition, the plots were smoothed by time averaging over 10 orbits. }
\label{figTimeEvo}
\end{figure}

\begin{figure}[h!]
\centering
\includegraphics[width=\columnwidth]{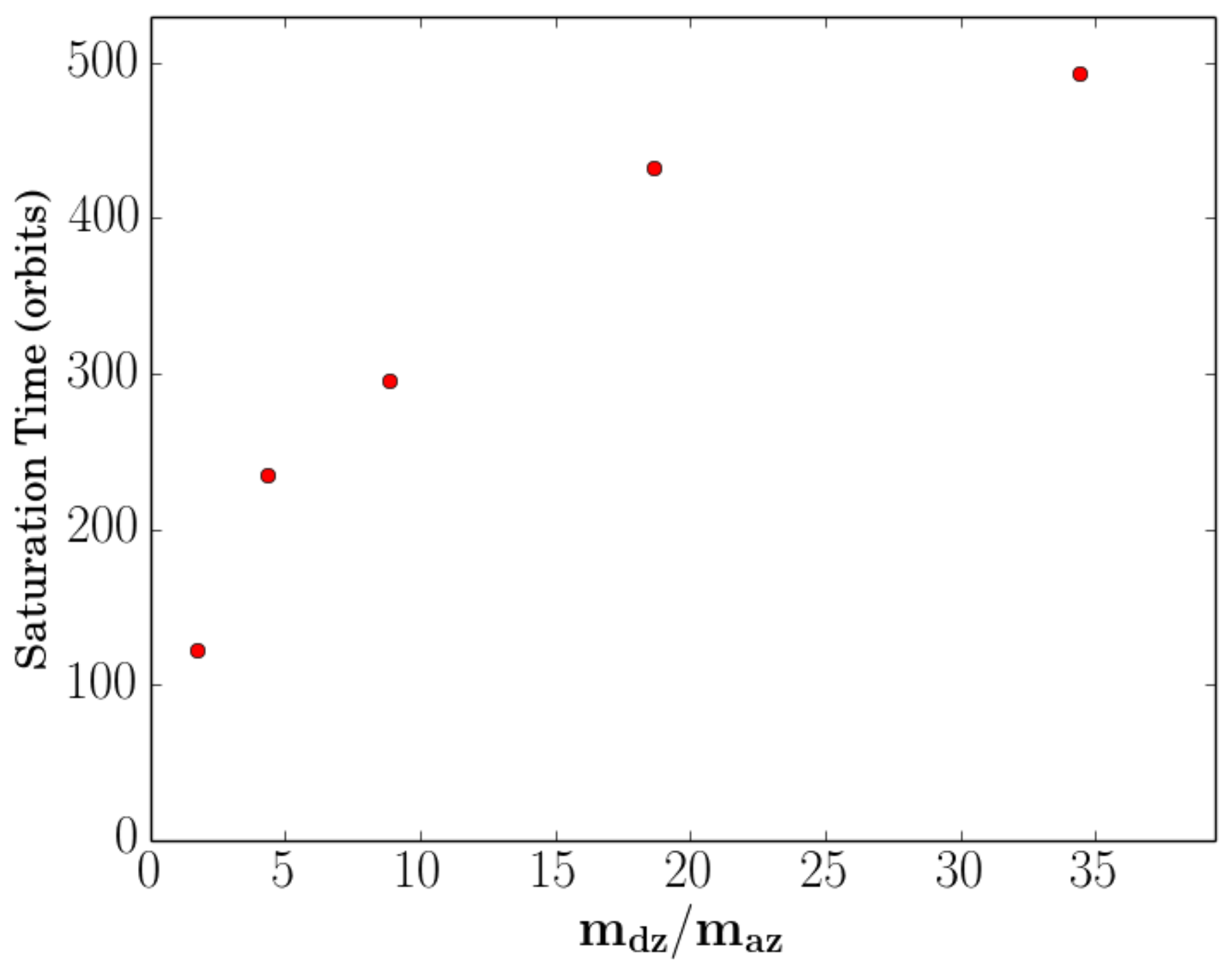}
\caption{The saturation time as a function of the ratio of the dead zone mass to the active zone mass. }
\label{figSatTime}
\end{figure}

Figure~\ref{figTimeEvo} shows the time evolution of the kinetic energy, Reynolds stress and Maxwell stress in the 
active and dead zone regions, for the run with a medium-sized dead zone. The stresses in the active zone saturate on a time scale of 10-20 orbits, consistent with previous studies of MRI-driven turbulence \cite[e.g.,][]{simon09}. The properties of turbulent  and non-turbulent fluid motions in the dead zone, conversely, evolve in a 
substantially different way. The dead zone Reynolds stress appears to initially saturate on a rapid time scale, but 
continuing the run reveals long ($\sim 10^2$ orbit) time scale, large amplitude fluctuations. The kinetic energy, 
meanwhile, grows essentially monotonically and saturates only after approximately 300 orbits of evolution. (The 
dead zone Maxwell stress also continues to evolve for a long time, but at such a low level as to be physically 
uninteresting.) As 
we will discuss later, the bulk of the kinetic energy is associated with non-turbulent fluid motions, and hence the 
kinetic energy evolves in a distinctly different manner to the Reynolds stress. To quantify how the saturation 
time scale depends on the dead zone thickness, we take the most conservative (slowest saturating) measure  
and define the saturation time to be when the average mid-plane kinetic energy did not increase by more than 
3 percent over 20 orbits. The saturation time increases as the mass of the dead zone increases relative to the 
active zone, as shown in Figure~\ref{figSatTime} and Table~\ref{tabResults}. This is qualitatively consistent with the 
expectation that energy and turbulence are being generated in the active layer and injected into the dead zone, so that a larger dead zone takes longer to saturate. However, the detailed evolution of the kinetic energy as a function of 
time (both for the dead zone shown in Figure~\ref{figTimeEvo} and the other runs) proved hard to fit with a 
simple quantitative model of energy injection plus damping. Clearly, though, some physically realistic dead zones --- for which the ratio of the active to dead mass might be as low as $10^{-3}$ --- would require very long simulations, with well-controlled numerical dissipation --- to model faithfully.

\subsection{Stresses and energies in the saturated state dead zone}
\begin{figure*}[t]
\centering
\includegraphics[width=\columnwidth]{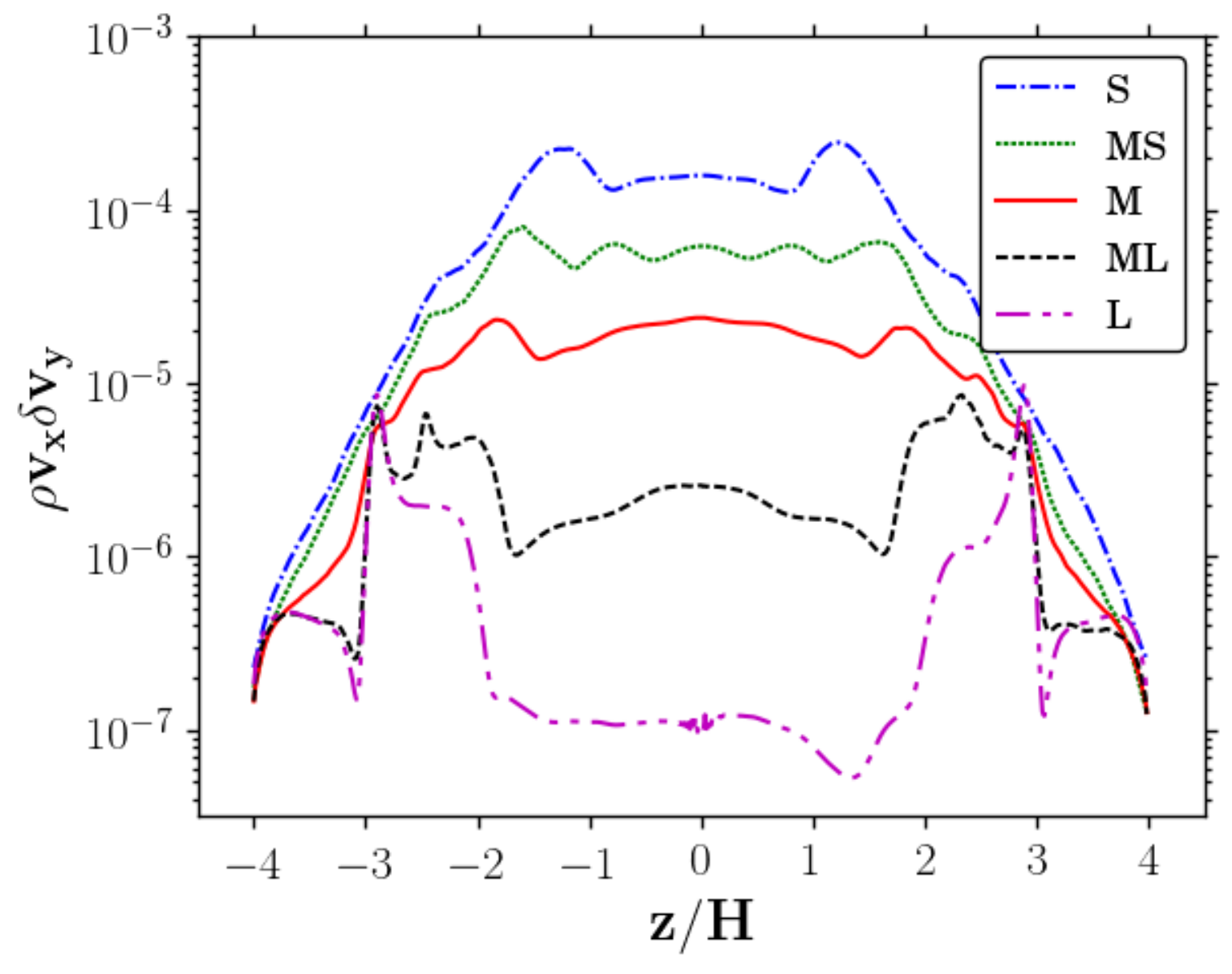}
\includegraphics[width=\columnwidth]{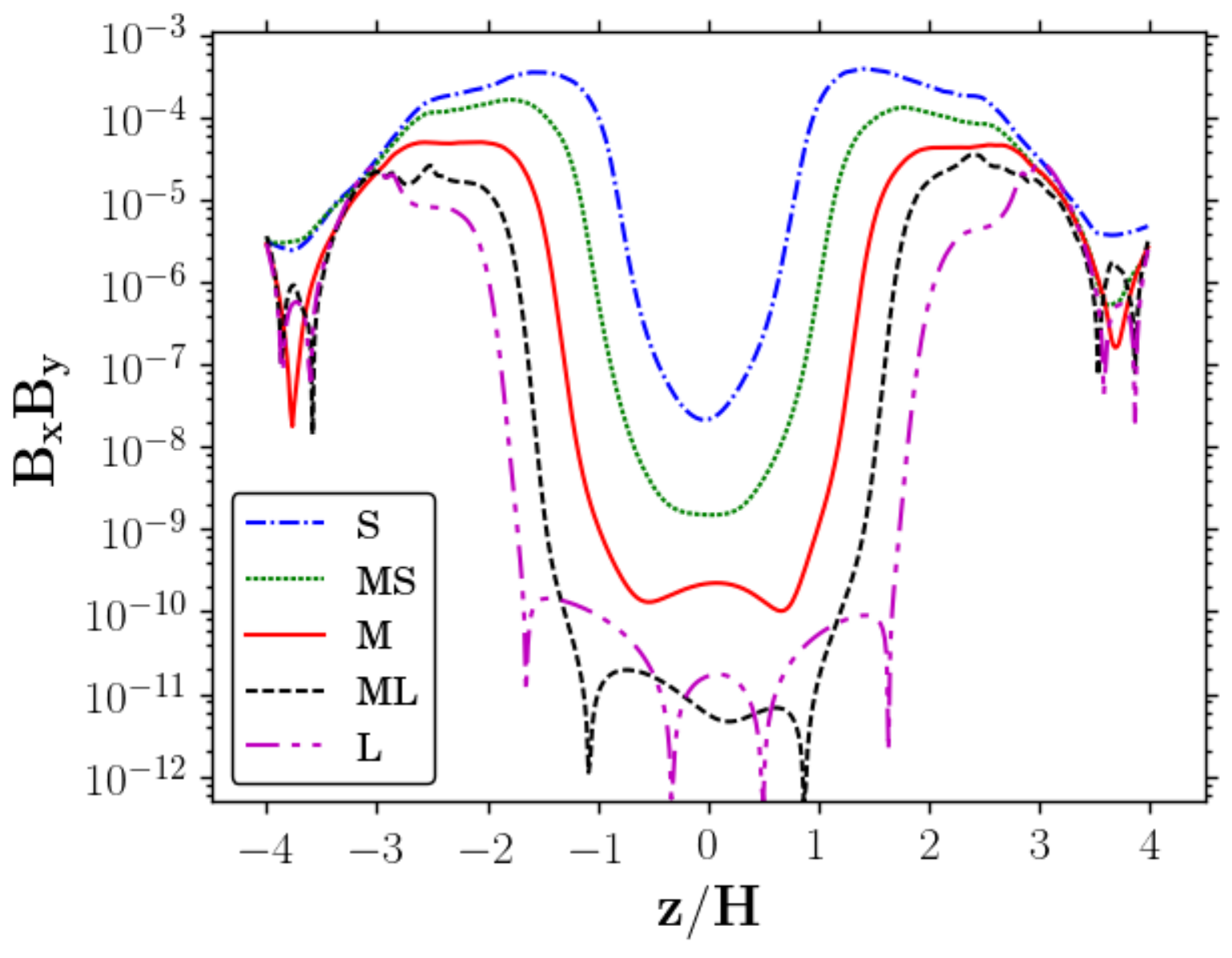}
\caption{Time averaged vertical profiles of the Reynolds stress (left) and Maxwell stress (right) for the different sized dead zones.  The sharp transition seen near $|z|=3$ in the Reynolds stress for the ML and L runs corresponds to where the r modes (discussed in detail in section 3.8) shut off.  The r modes have a larger spatial extent in the ML and L runs due to the larger dead zone.  Therefore the location of the transition feature happens in a region of lower Reynolds stress, making the features most prominent in these runs. }
\label{figProfiles}
\end{figure*}

Having defined the saturation time for the different dead zone sizes, we compute time-averaged vertical profiles 
of the Reynolds and Maxwell stresses using data from the remaining portion of the runs. These stress profiles are plotted in Figure~\ref{figProfiles}. The mid-plane Maxwell and 
Reynolds stresses both decline as the thickness of the dead zone increases, but the Maxwell stress is always 
negligible. Only the Reynolds stress is interesting for physical purposes.

\begin{figure}[h]
\centering
\includegraphics[width=\columnwidth]{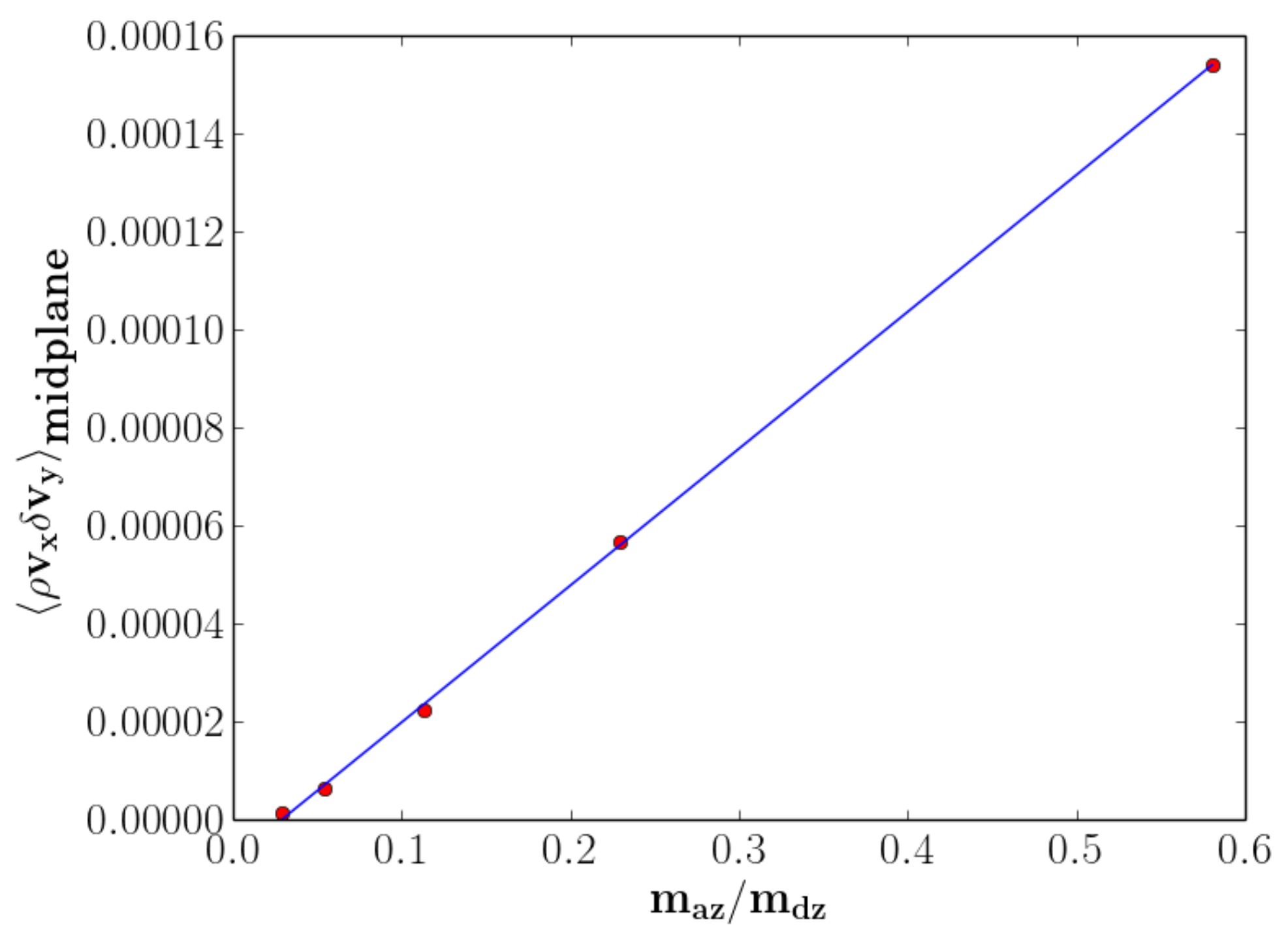}
\caption{The average Reynolds stress near the mid-plane as a function of the ratio of the active zone mass to the dead zone mass.  Note that the mass ratio on this x-axis is inverted relative to figures 6 and 11 to show the linear relationship.  }
\label{figReynoldsPlot}
\end{figure}

If Reynolds stress in the dead zone derives exclusively as a consequence of energy transported there from the 
active zone, then its magnitude ought to be set by an equilibrium between energy injection from the active zone 
and damping (which might be turbulent dissipation in the dead zone, or leakage of dead zone energy back 
into the active zone). We therefore expect a scaling of the mid-plane Reynolds stress with the ratio of the mass 
in the active zone to that in the dead zone. The derived scaling is shown in Figure~\ref{figReynoldsPlot}. A linear 
fit to the $\langle \rho v_x \delta v_y \rangle$-$m_{\rm az} / m_{\rm dz}$ relation provides a good description of the numerical results at the disk mid-plane. As can be seen in Figure~\ref{figProfiles}, however, the mid-plane stress becomes 
extremely small for the thickest dead zone, and this is reflected in the linear fit predicting that the mid-plane 
stress goes to zero at finite dead zone thickness. We do not expect this to be strictly correct, but the results 
suggest that Ohmic dead zones have physically negligible mid-plane stresses once the active zone column  
becomes less than or of the order of 1\% of the total disk column.

\subsection{Transport Efficiency}

The ratio of Reynolds stress to kinetic energy (KE) is indicative of the efficiency of angular momentum transport.  
The ratio is defined as,

\begin{equation}
\frac{\langle \text{reynolds stress} \rangle}{\langle \text{KE} \rangle} = \frac{\langle \rho v_r  v_\phi \rangle}{\langle \rho v^2 /2 \rangle}                      
\end{equation}

\noindent where the brackets indicate an average over a given spatial volume.  Table~\ref{tabResults} contains this efficiency diagnostic for the different runs, time averaged in the saturated state. The dead zone spatial average is taken from from $|z/H|$ $= 0$ to $0.5$, while the active zone is from $|z/H|$ $=2$ to $3.3$.  These limits were maintained across the runs to compare like quantities.

The efficiency in the active zone is about $0.17$, independent of the size of the dead zone, while the efficiency in the dead zone scales strongly with the size of the dead zone.  The (hydrodynamic) transport efficiency for the medium sized dead zone is about an order of magnitude less than in the active region, where ideal MHD conditions apply. There is a substantial further decline for the larger dead zones. The large kinetic energy content of the dead zone is thus largely irrelevant as far as angular momentum transport goes, implying some combination of (a) non-turbulent motions and (b) turbulence that at the mid-plane is inefficient at transport.

\begin{table*}
\centering
 \begin{tabular}{|l | c | c | c | c | c | c |} 
 \hline
 DZ Size & Sat. Time (orbits) & Time Averages & Reynolds $\frac{active}{dead}$ & KE $\frac{active}{dead}$ & Trans. Eff. Dead & Trans. Eff. Active \\  
 \hline
 S & 120 & 150-230 & 40.1 & 11.6 & 0.048 & 0.17\\
 \hline
  MS & 230 & 230-300 & 67.5 & 10.2 & 0.023 & 0.15 \\
 \hline
  M & 290 & 250-450 & 114 & 13.8 & 0.02 & 0.16 \\
 \hline
 ML & 430 & 400-700 & 2250 & 51 & 0.0098 & 0.18 \\
 \hline
 L & 490 & 500-800 & 4550 & 106 & 0.0041 & 0.18\\
 \hline
\end{tabular}
\caption{Summary of time and volume averaged values in the steady state.}
\label{tabResults}
\end{table*}

\subsection{Density Perturbations}
The magnitude of density fluctuations is of interest, especially in the mid-plane where stochastic 
gravitational forces from such fluctuations excite the velocity dispersion of planetesimals 
\citep{ida08,gressel11,okuzumi13}. We define the amplitude of density perturbations as,

\begin{equation}
\frac{\delta\rho}{\rho}(x,y,z,t) = \frac{\rho(x,y,z,t)-\langle \rho \rangle_{xy}(z,t)}{\langle \rho \rangle_{xy}(z,t)}.                  
\end{equation}

\noindent The time and volume averaged density perturbation is,

\begin{equation}
\frac{\delta\rho}{\rho}(z) = \overline{\langle |\frac{\delta\rho}{\rho}(x,y,z,t)|\rangle_{xy}},                  
\end{equation}

\noindent where the over-bar represents a time average and the brackets represent a volume average over the given dimensions.

\begin{figure}[h]
\centering
\includegraphics[width=\columnwidth]{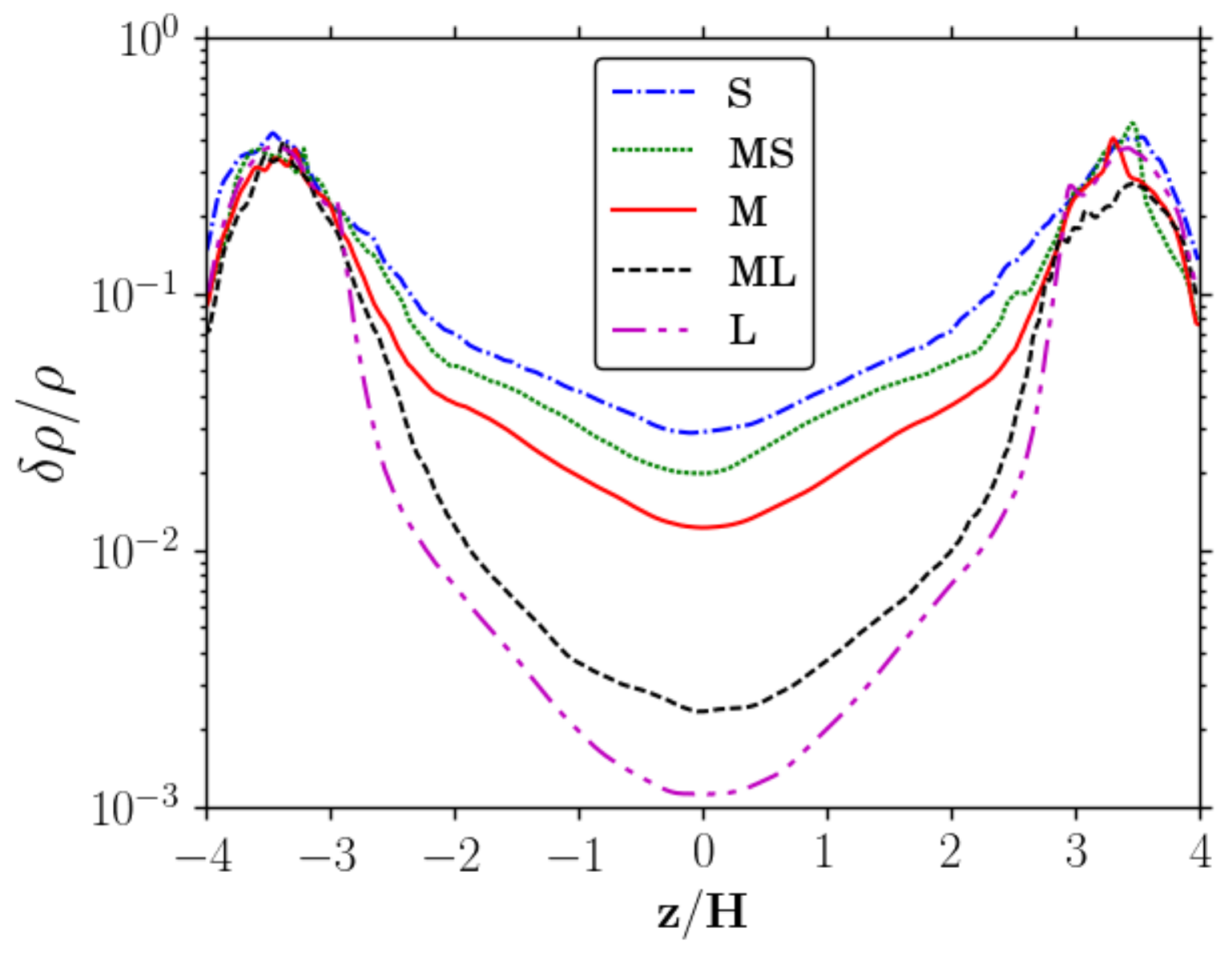}
\caption{The density perturbation profiles for the 5 runs, averaged over the saturated state.}
\label{figDensPertProfiles}
\end{figure}

The vertical profile of the density perturbations in the saturated state can be seen in Figure~\ref{figDensPertProfiles} for each different run.  The perturbation in the active layer is relatively consistent across the different runs, peaking at about $\simeq 0.25$ (i.e. moderately compressible).  However, the perturbation at the mid-plane decreases as the dead zone size increases, being as high as 0.02 for the smallest dead zone and $3 \times 10^{-3}$ for the largest.  This diagnostic again reveals that, while there is a large amount of kinetic energy in the dead zone, the fluid motions are essentially incompressible and are not turbulent --- even hydrodynamically --- in the same way as the active zone.

\subsection{Accretion Rates}

\begin{figure}[h]
\centering
\includegraphics[width=\columnwidth]{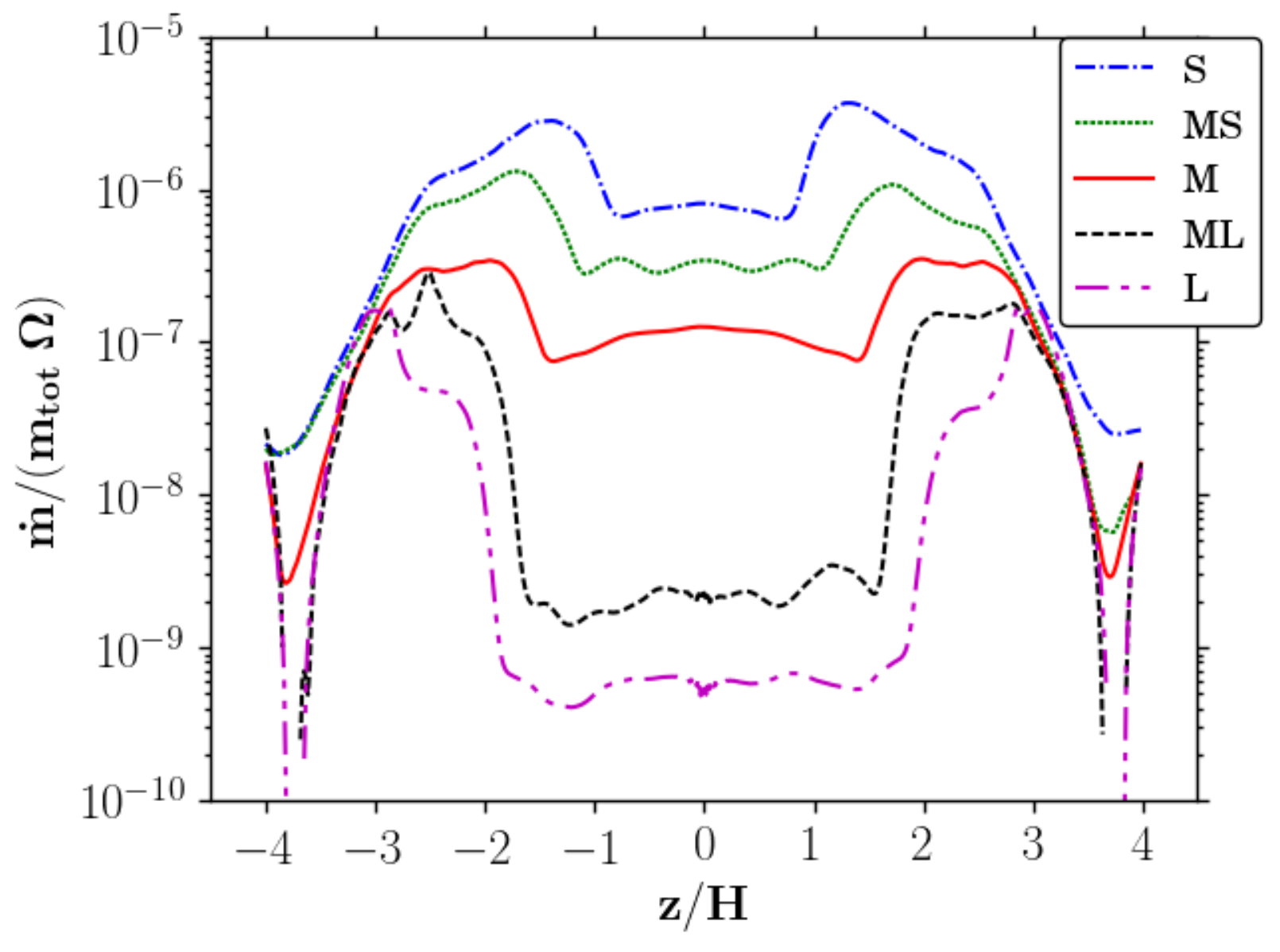}
\caption{The non-dimensional accretion rate as a function of $z$ for the different dead zone sizes, averaged over the saturated state.}
\label{figAccretionProfiles}
\end{figure}

\begin{figure}[h]
\centering
\includegraphics[width=\columnwidth]{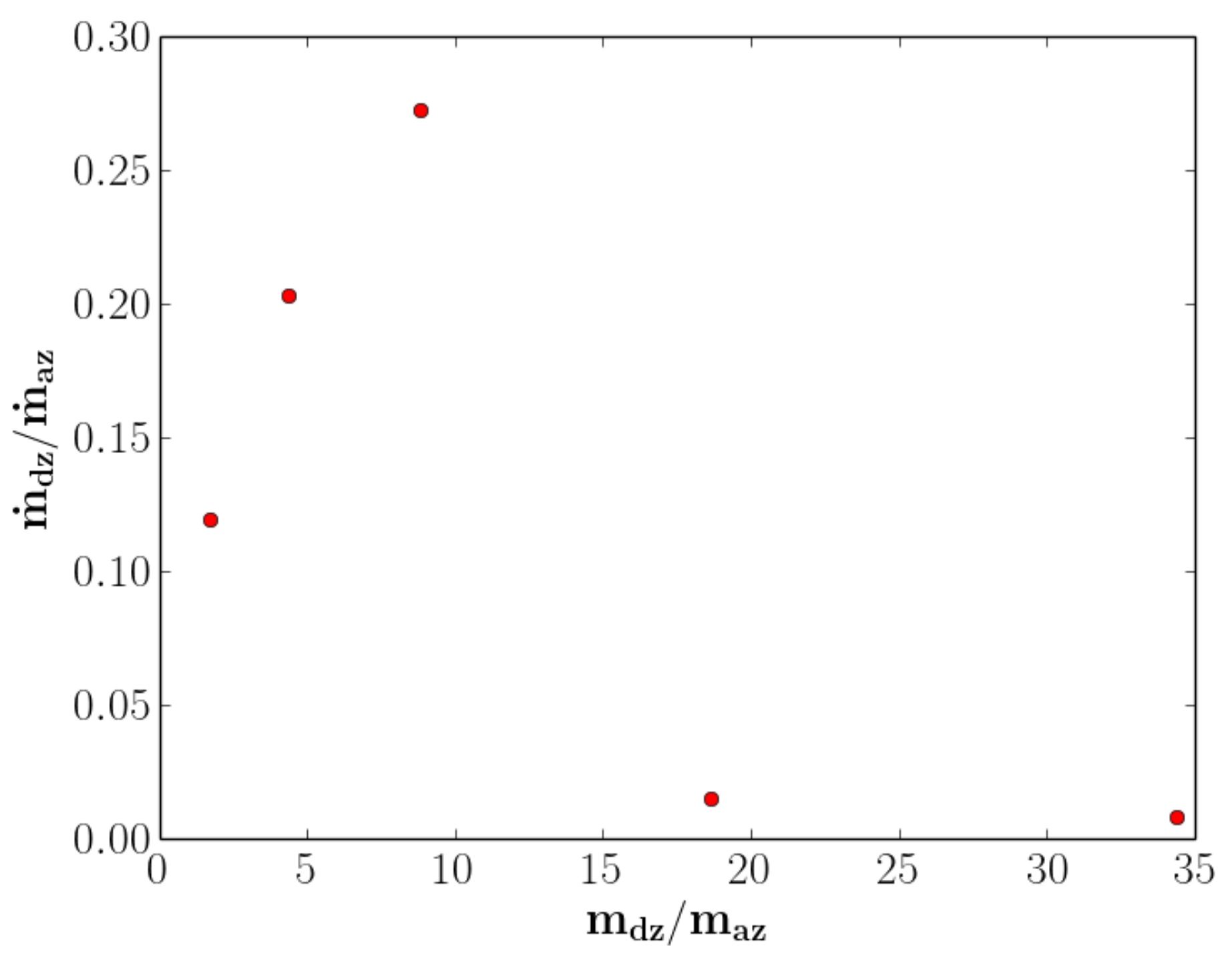}
\caption{The ratio of the accretion rate in the dead zone to the accretion rate in the active zone as a function of the mass ratio of the two zones.  The accretion rate was averaged over the region defined to be the dead zone in table \ref{tabSizes}, which changes from run to run.  Time averaging was done over the saturated state.  }
\label{figAccretionMassRelation}
\end{figure}

Disks with dead zones are potentially subject to outbursts as the dead zone gains mass
that can rapidly accrete once mid-plane temperatures are high enough
for the MRI to set in \citep{armitage01,zhu09,martin11}. These outbursts can be avoided if the dead zone is sufficiently active, such as
in the steady state disk models of \cite{terquem08}.
However, such models generally require that the mass accretion rate
through the dead zone be larger than in the active region.
The simulations provide a means of estimating the accretion rates
through the active and dead 
zones by performing integrals $\propto \int \rho \alpha(z) dz$.
The dimensionless accretion rate from a single, constant $z$ plane is calculated as follows,

\begin{equation}
\label{mdot}
\frac{\dot{m}(z)}{m_{\text{tot}}\Omega} = \frac{3\pi c_s^2}{m_{\text{tot}}\Omega^2} \langle\rho\rangle_{xy} \bigg(\frac{\langle \rho v_x \delta v_y - B_x B_y\rangle_{xy}}{\langle P \rangle_{xy}}\bigg)\Delta z,             
\end{equation}    

\noindent where $m_{\text{tot}}$ is the total mass of the shearing box.  In Figure~\ref{figAccretionProfiles}, we plot this accretion rate profile for each run.
 We then integrate Equation~\ref{mdot} over $z$ to calculate the ratio of the mass accretion rate through the dead zone
to the mass accretion in the active zone --- parameter $f$ in equation 11
of \cite{martin14} --- as a function of the ratio of the dead to active zone masses. This ratio is shown in Figure~\ref{figAccretionMassRelation}.
The result is a non-monotic relationship in which $f$ is less than unity for all
plotted cases.
For the three smallest dead zones, the dominant effect is that adding dead zone mass increases $f$ simply because there is more mass present in the dead zone to accrete.  However, for the largest two dead zones, the dominant effect is that the stress is very low, which causes $f$ to be very small. Since $f$ is not large, we expect then that outbursts
will generally occur (see Fig. 5 of \cite{martin14}).

\subsection{Structure of the Turbulence}
\begin{figure*}[t]
\centering
\includegraphics[height=34mm, width=40mm]{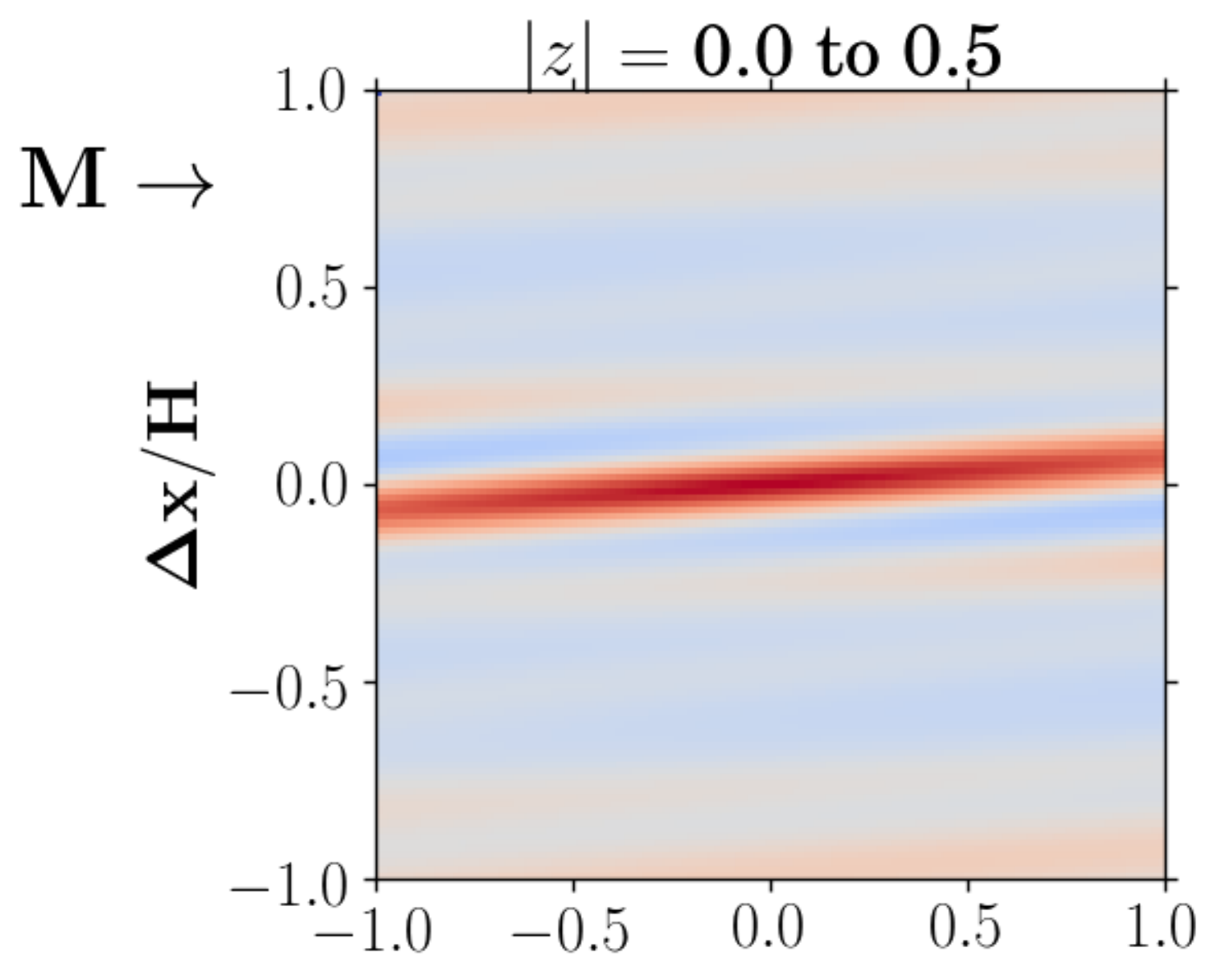}
\includegraphics[height=34mm, width=33mm]{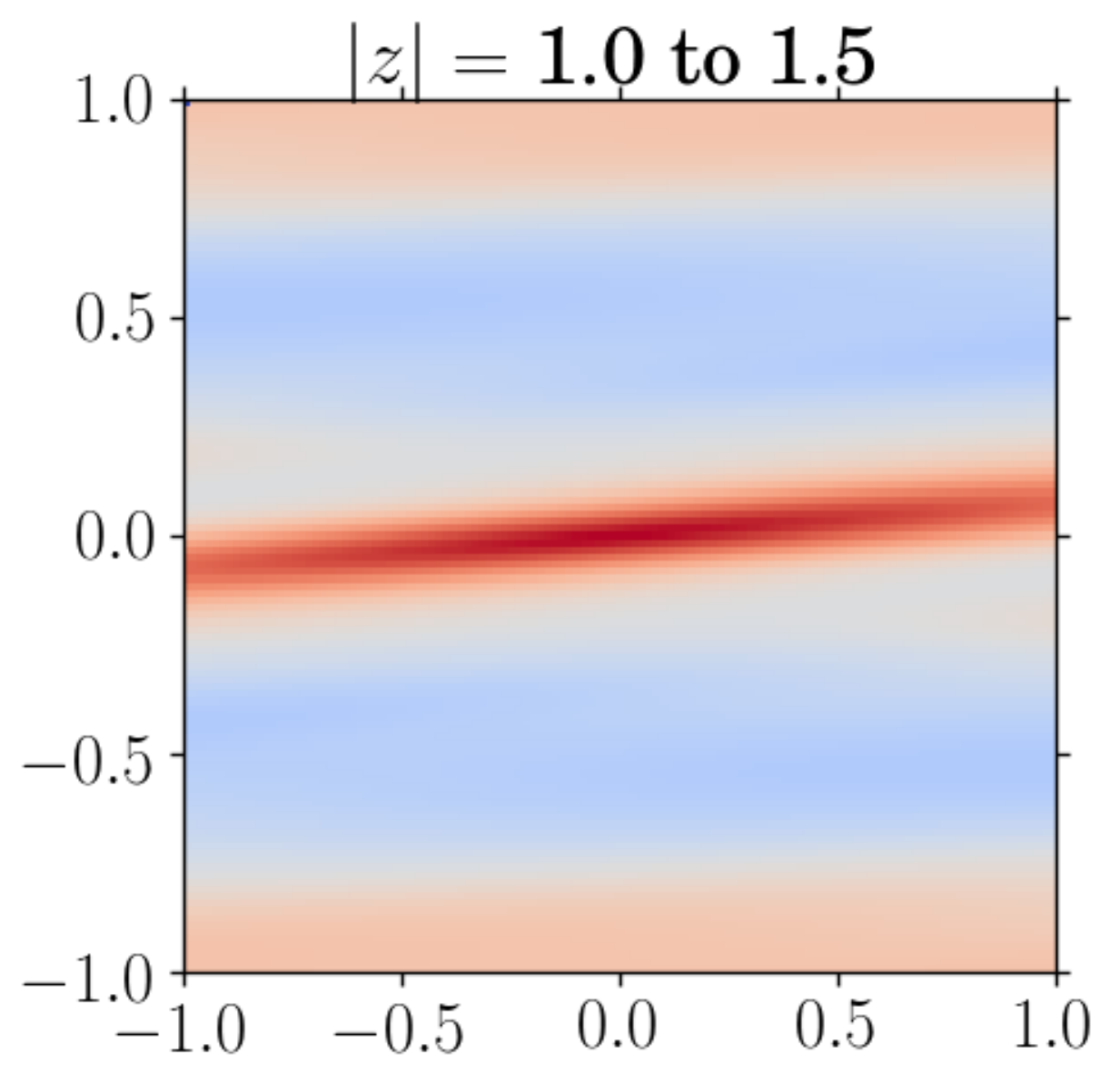}
\includegraphics[height=34mm, width=33mm]{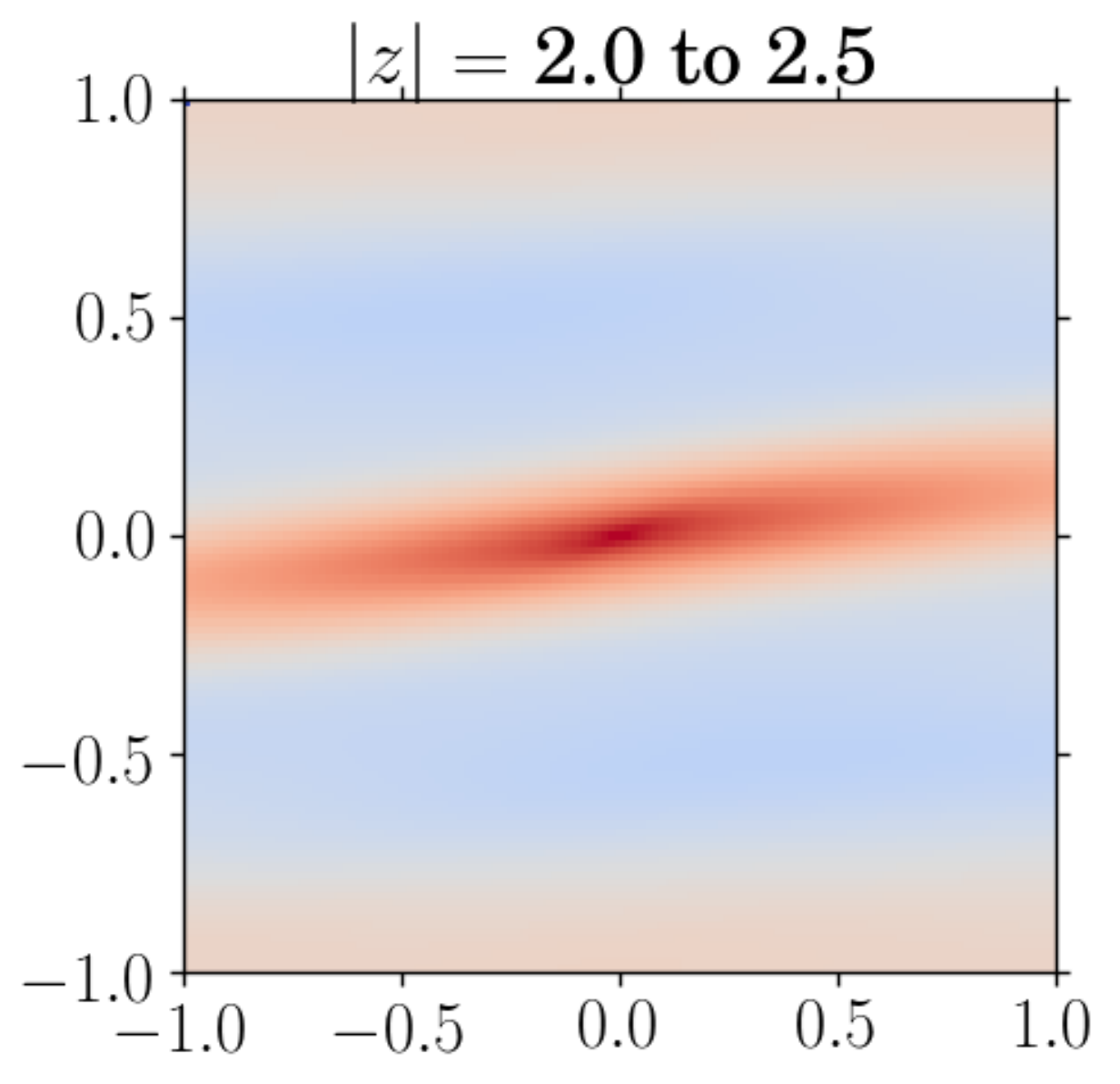}
\includegraphics[height=34mm, width=40mm]{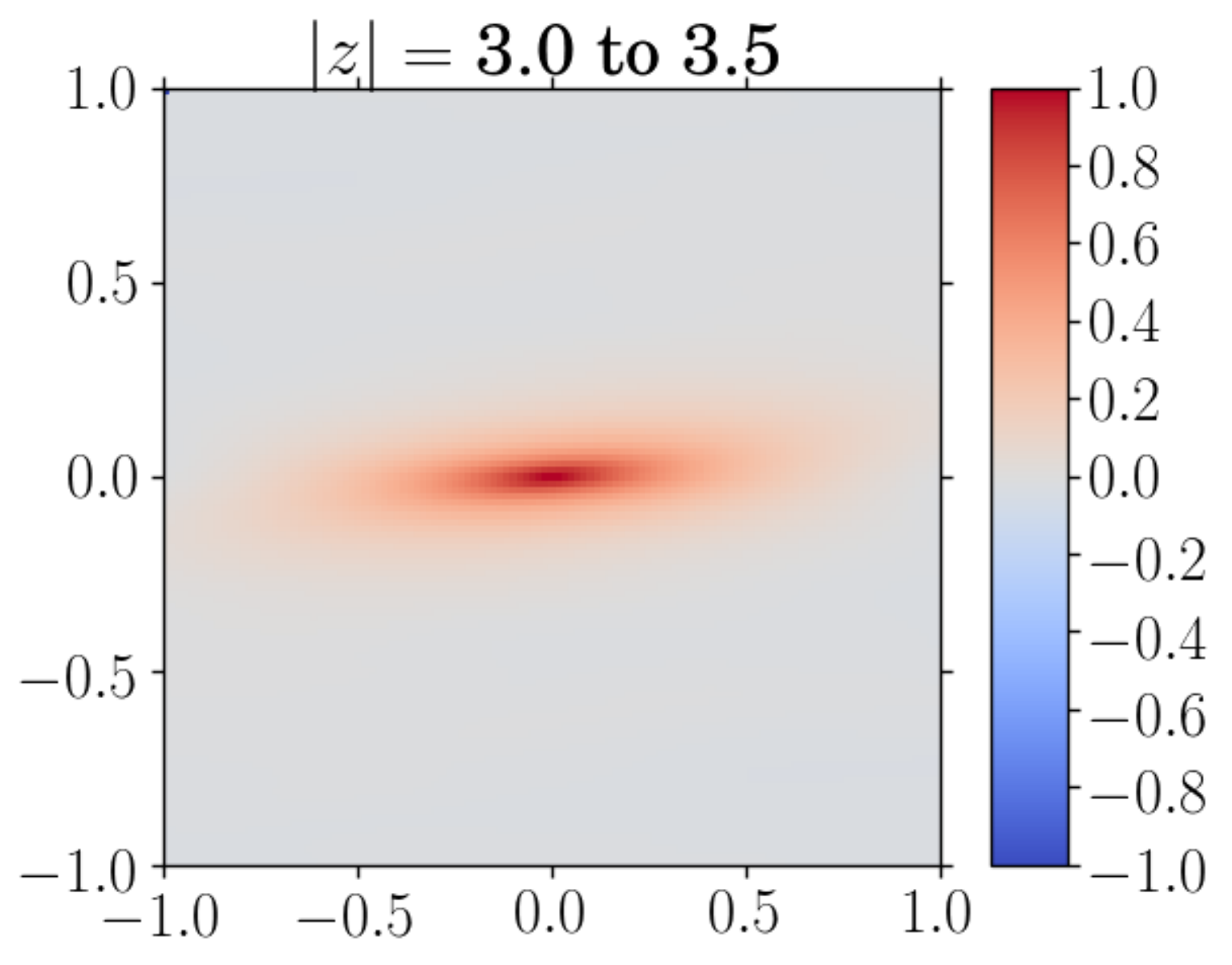}
\includegraphics[height=34mm, width=40mm]{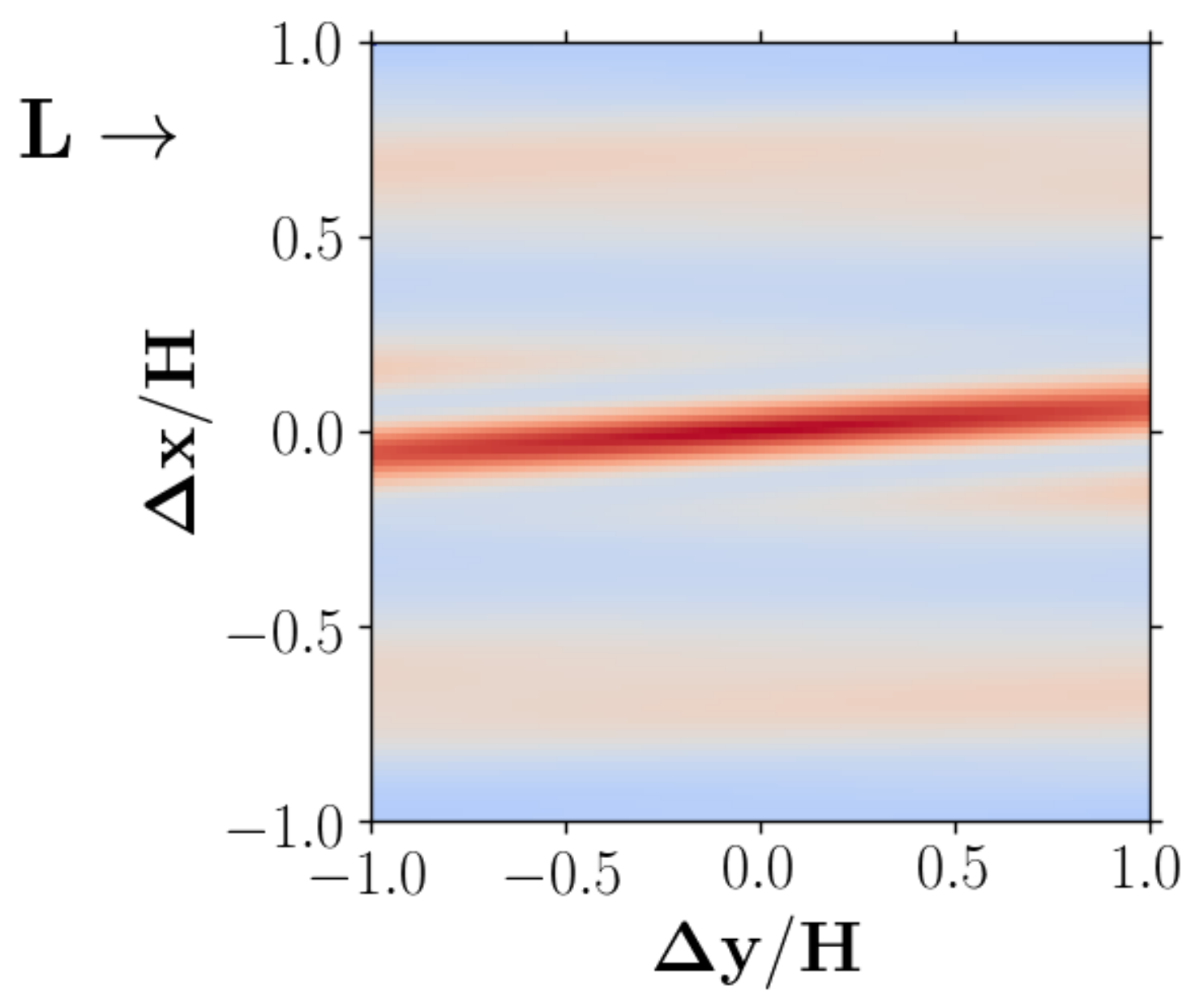}
\includegraphics[height=34mm, width=33mm]{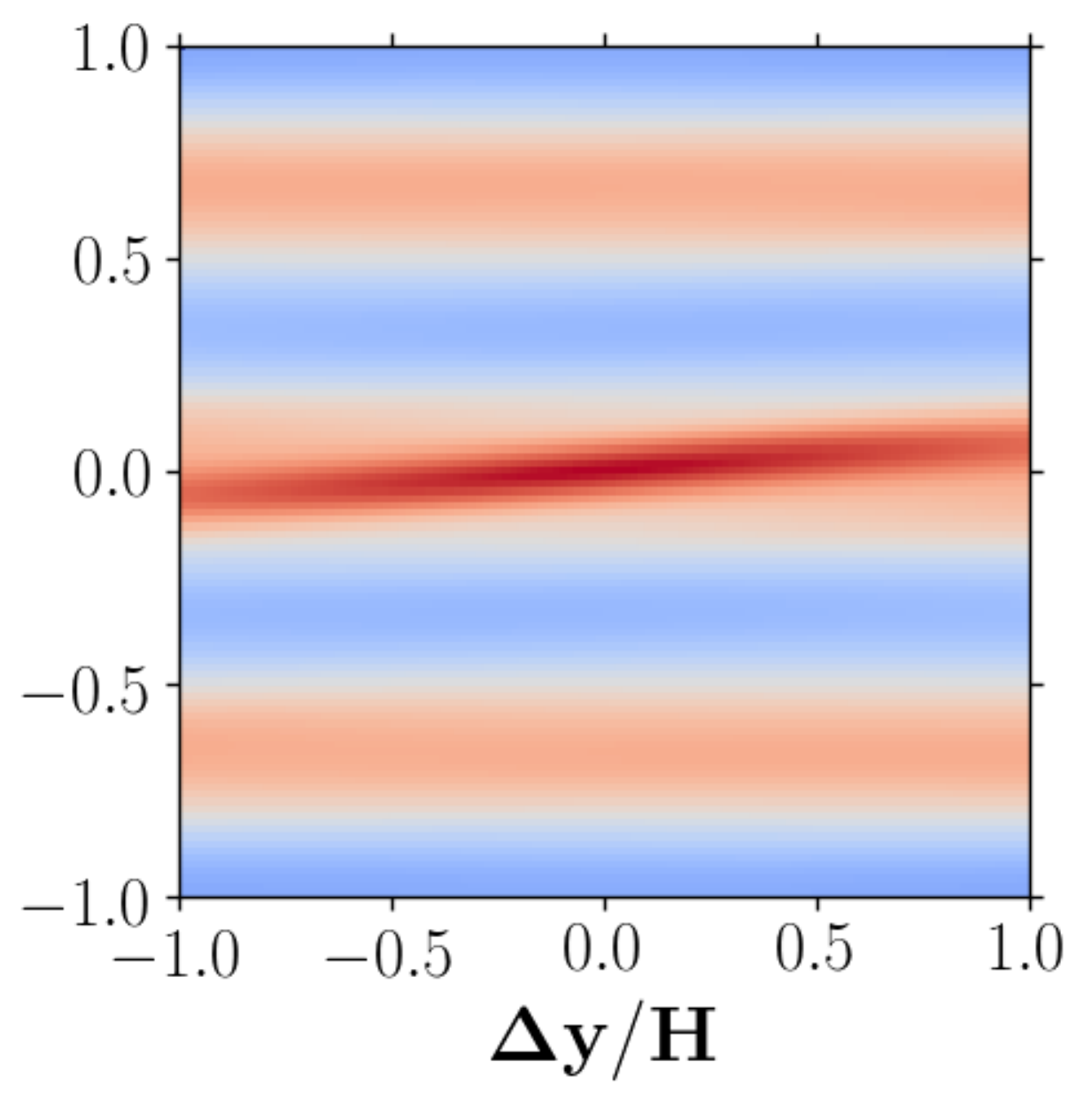}
\includegraphics[height=34mm, width=33mm]{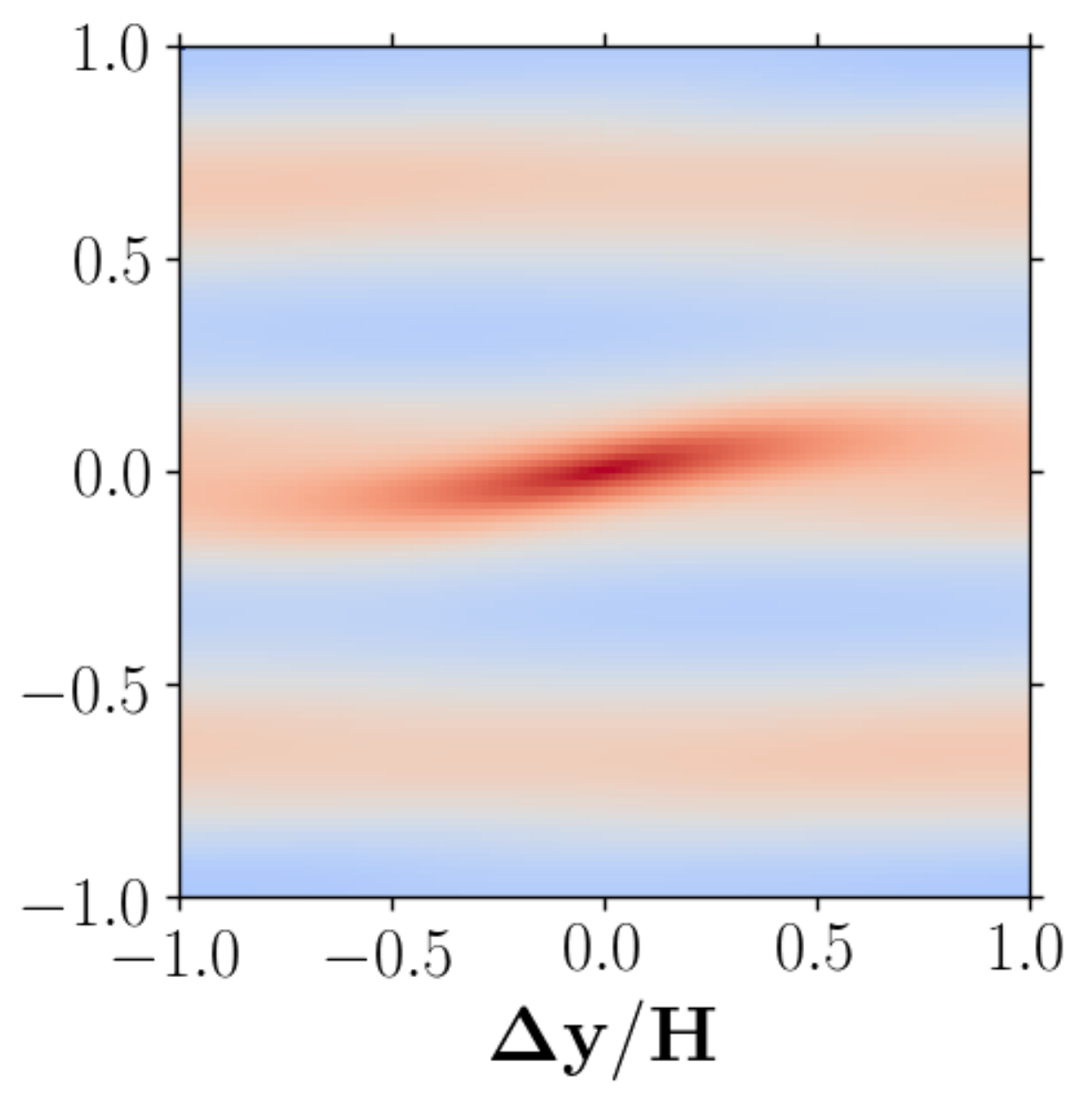}
\includegraphics[height=34mm, width=40mm]{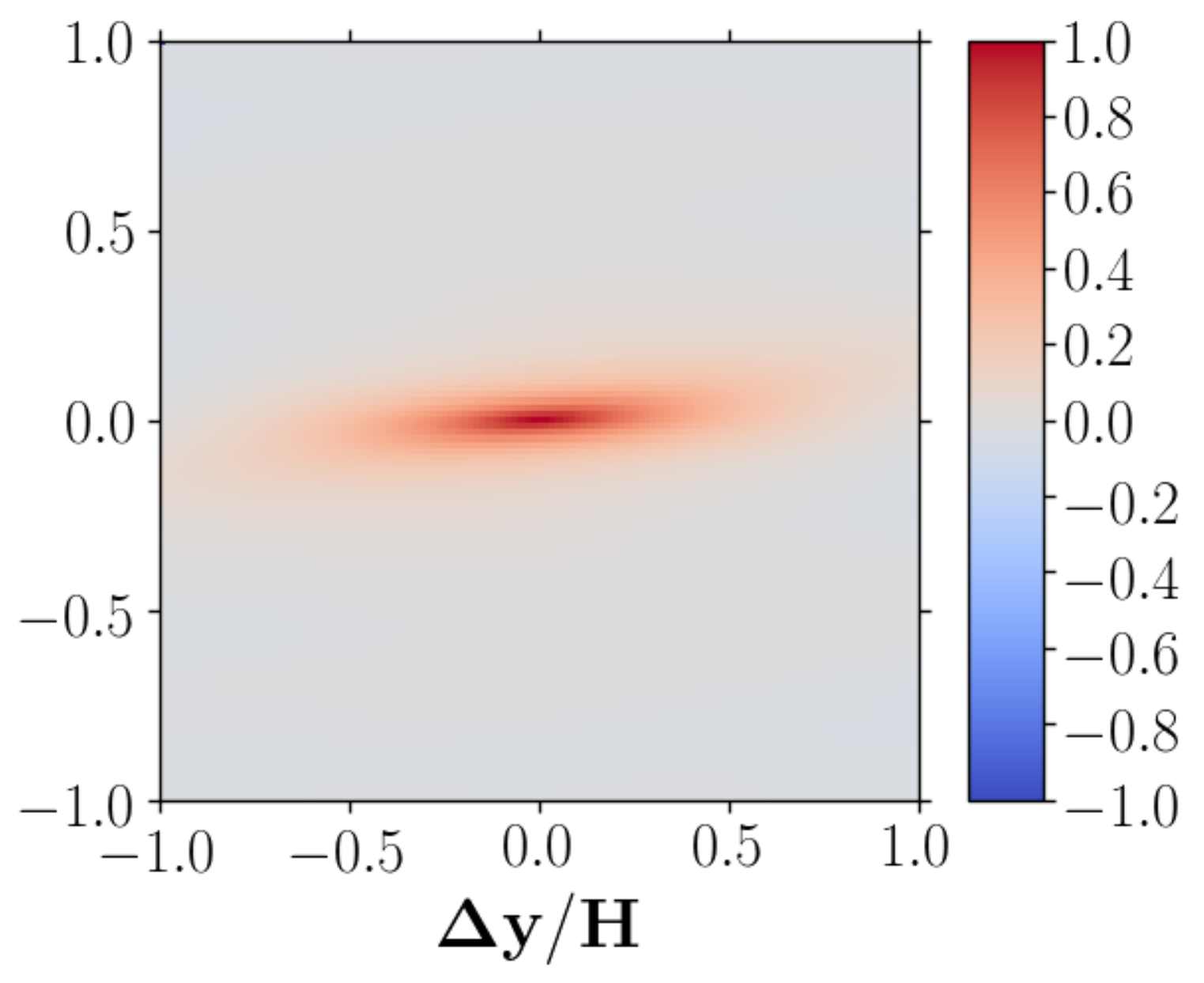}
\caption{Autocorrelation functions for the density perturbation for the medium dead zone (top) and large dead zone (bottom).  Moving from left to right, the ACF is calculated using planes $|z|=$ 0 to 0.5, 1.0 to 1.5, 2.0 to 2.5 and 3.0 to 3.5.  Each plot is $2H$ on a side.  The plots on the left show that perturbations near the mid-plane get sheared out by the background flow, while perturbations in the upper layer are smaller and less aligned with the y-axis.  }
\label{figACFdensPert}
\end{figure*}

\begin{figure*}[t]
\centering
\includegraphics[height=34mm, width=40mm]{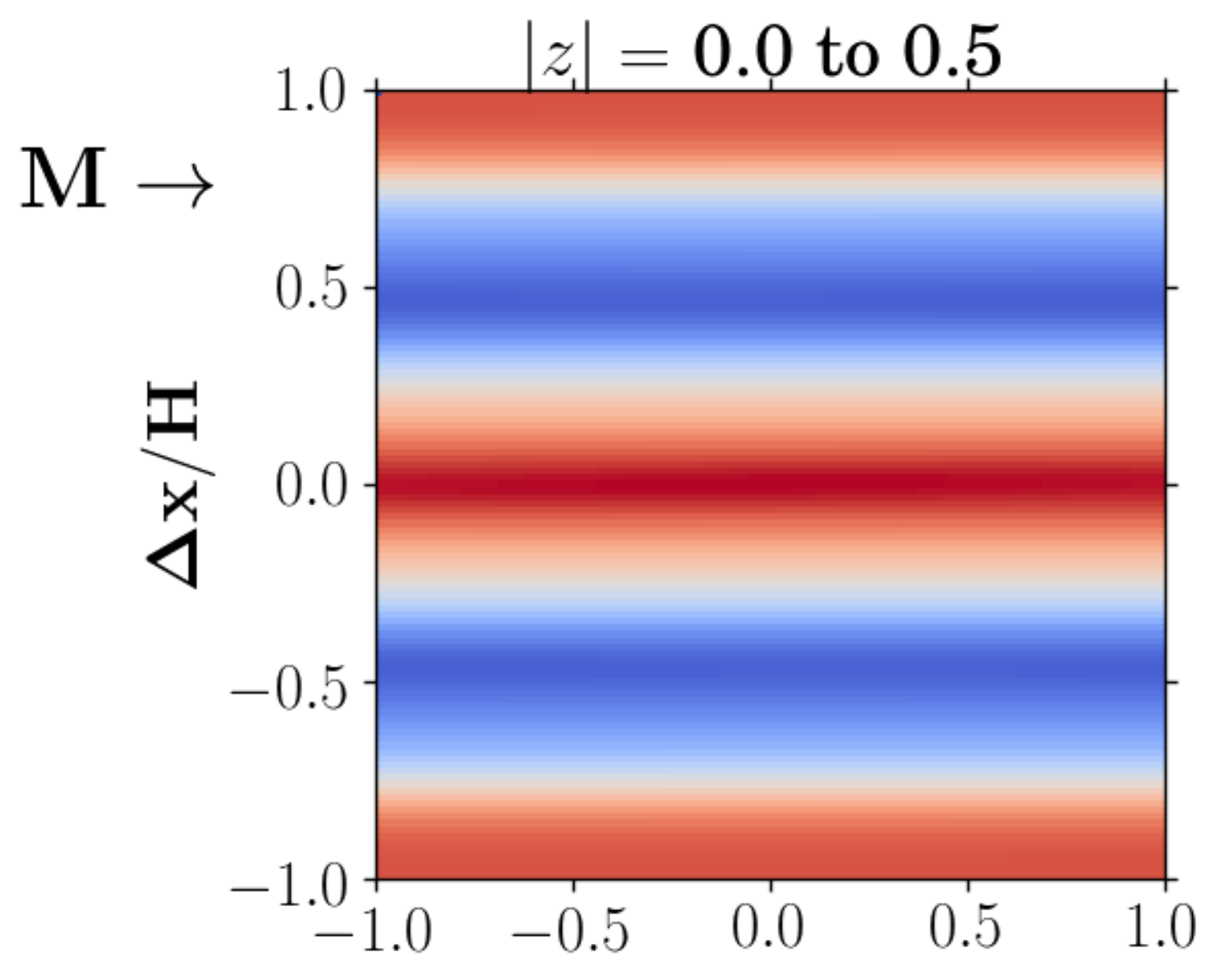}
\includegraphics[height=34mm, width=33mm]{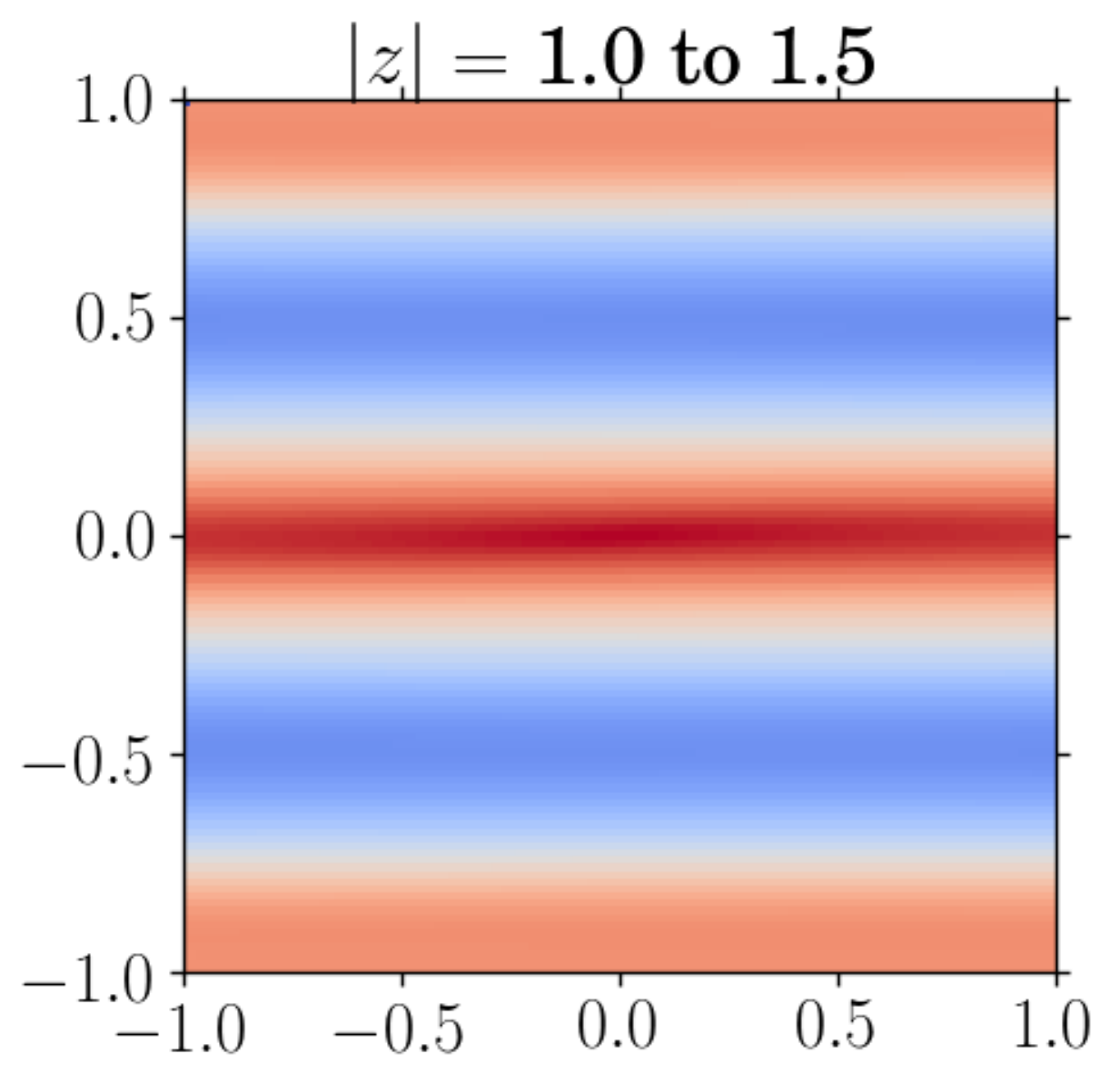}
\includegraphics[height=34mm, width=33mm]{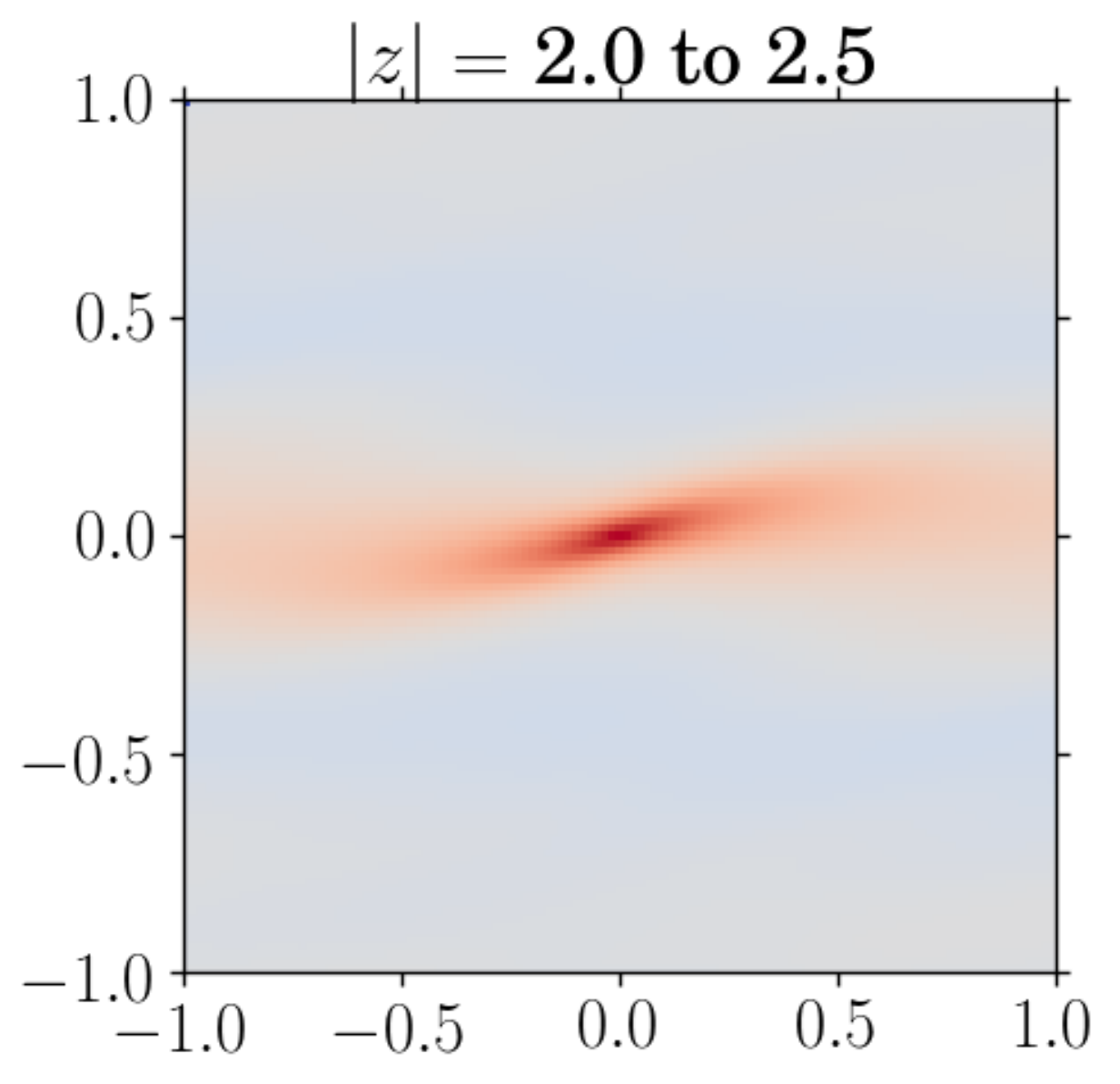}
\includegraphics[height=34mm, width=40mm]{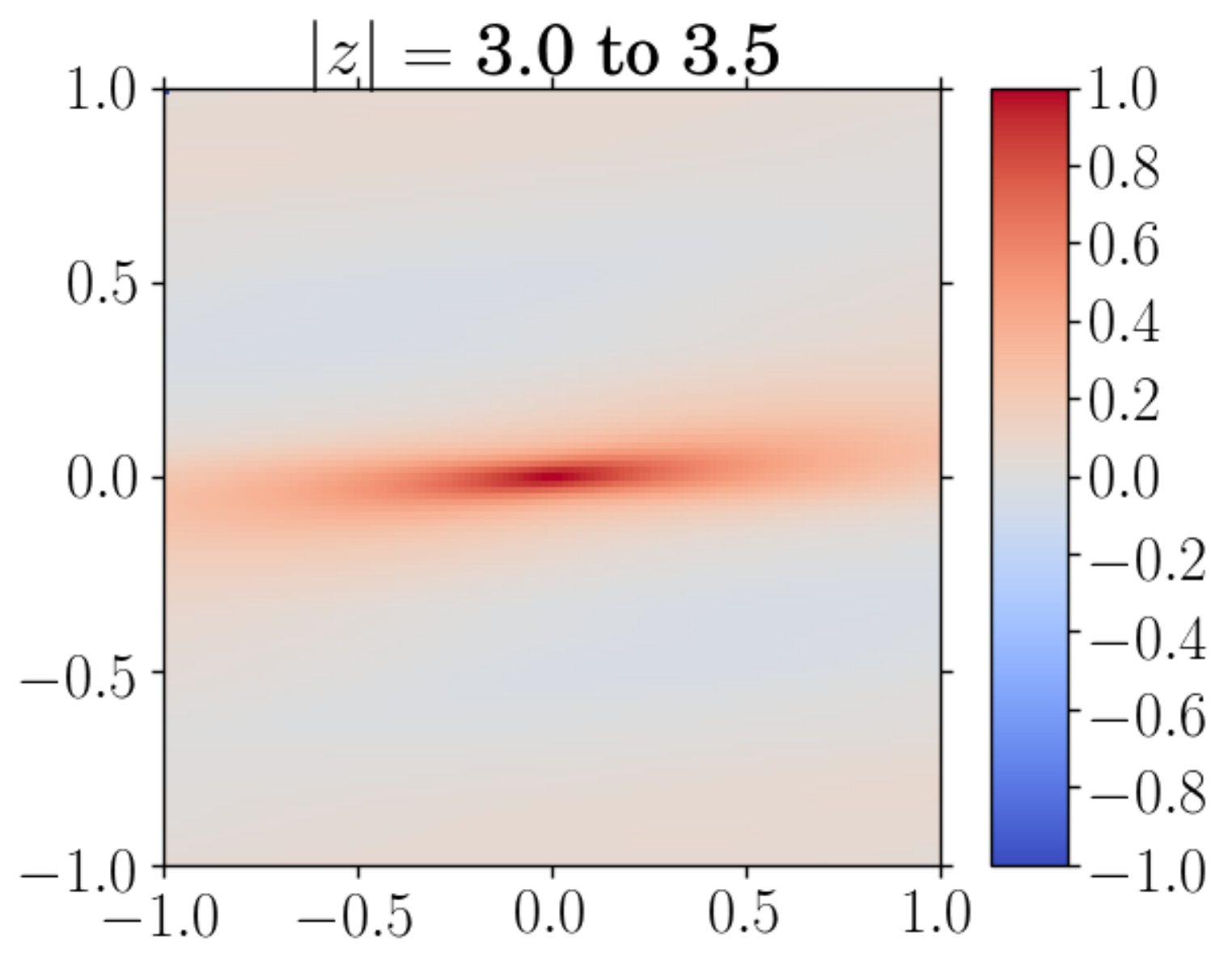}
\includegraphics[height=34mm, width=40mm]{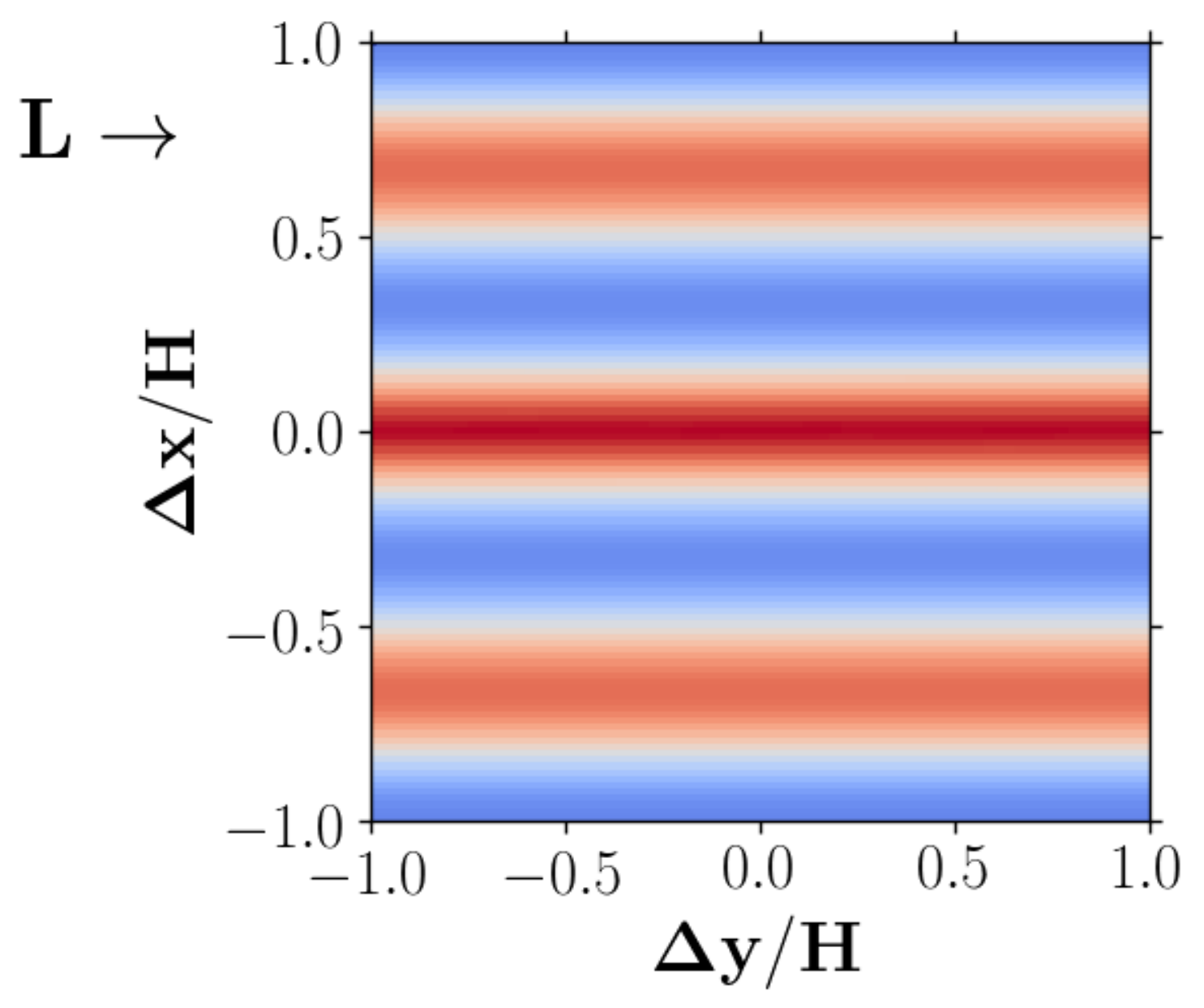}
\includegraphics[height=34mm, width=33mm]{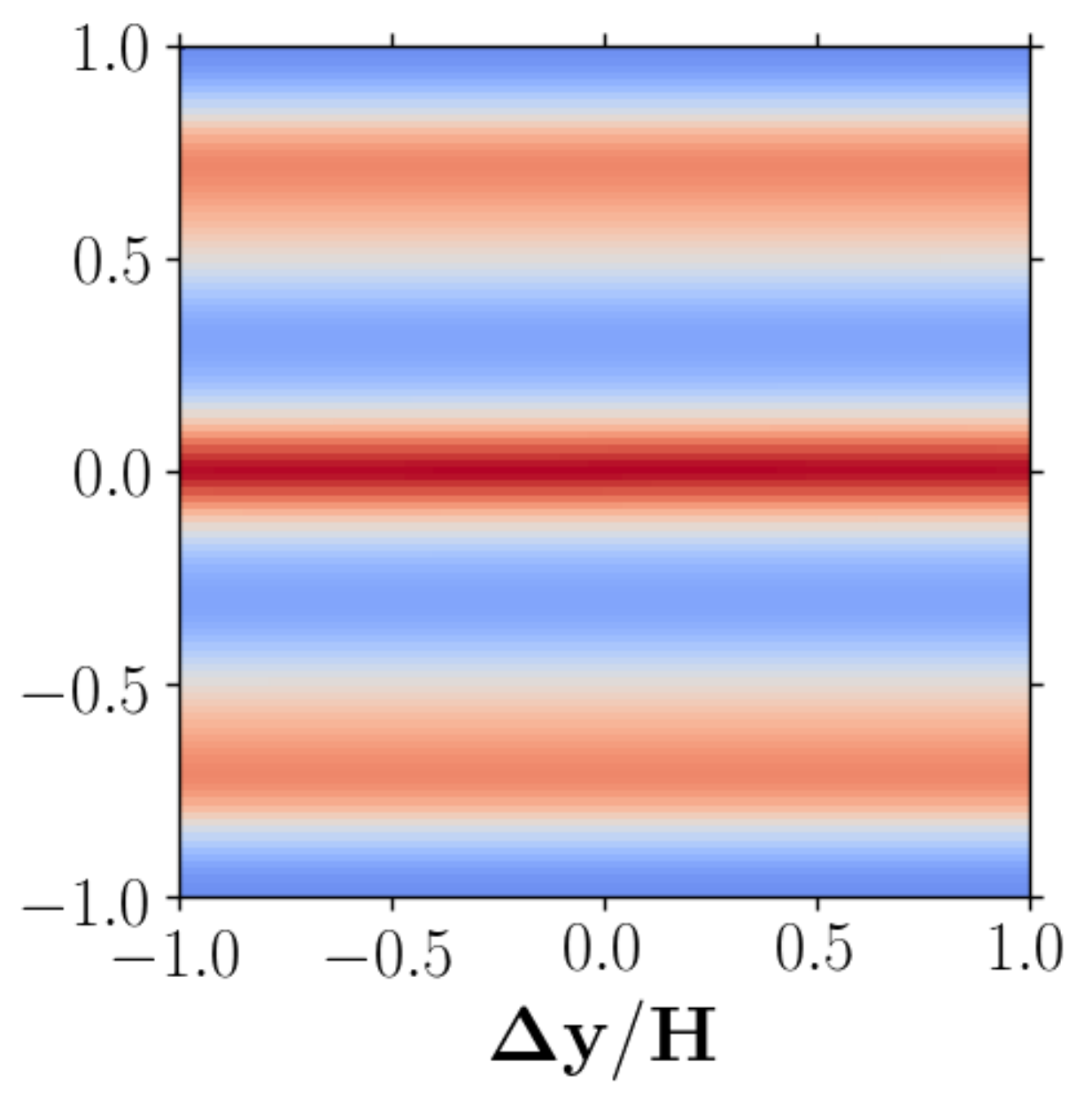}
\includegraphics[height=34mm, width=33mm]{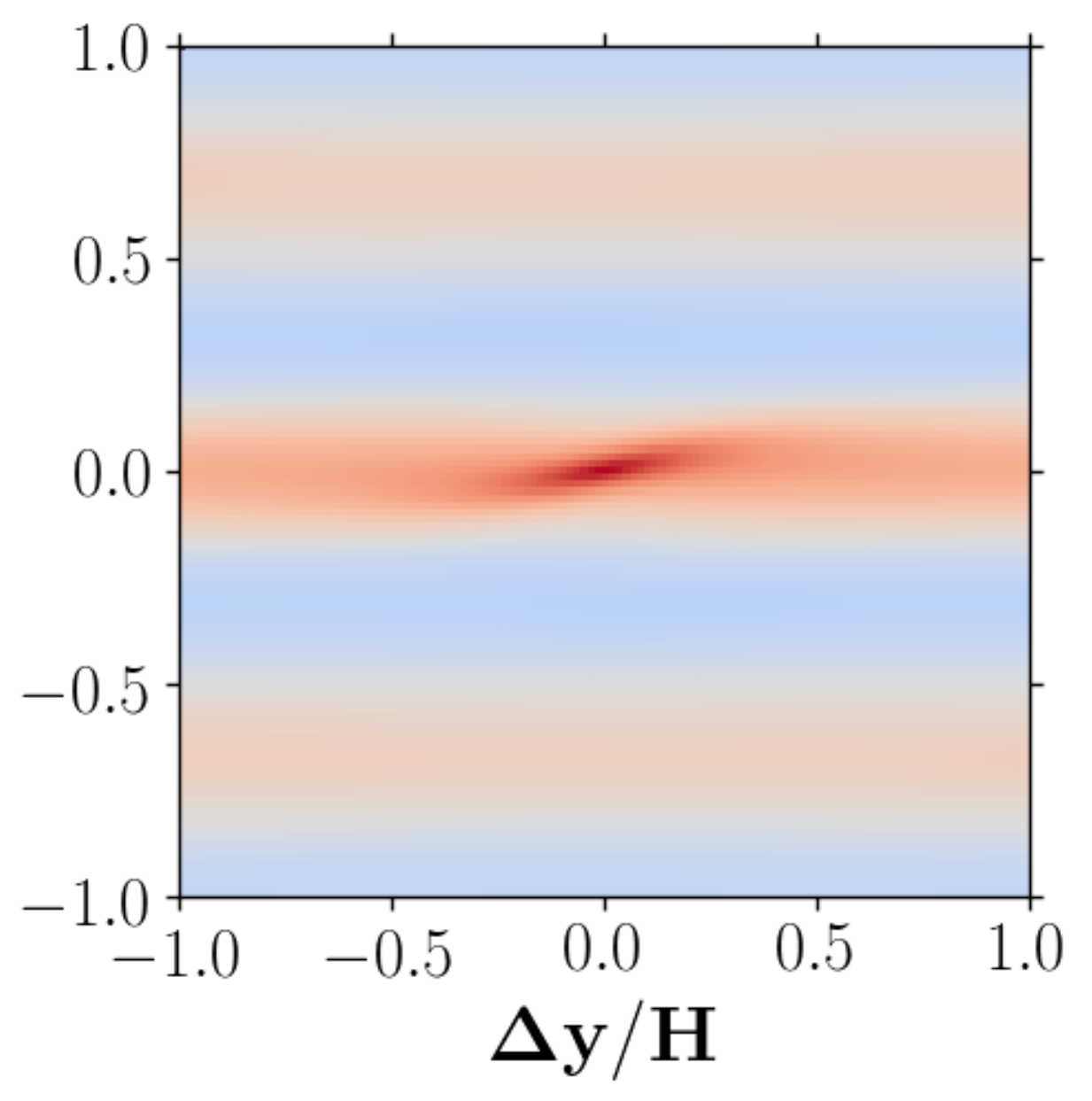}
\includegraphics[height=34mm, width=40mm]{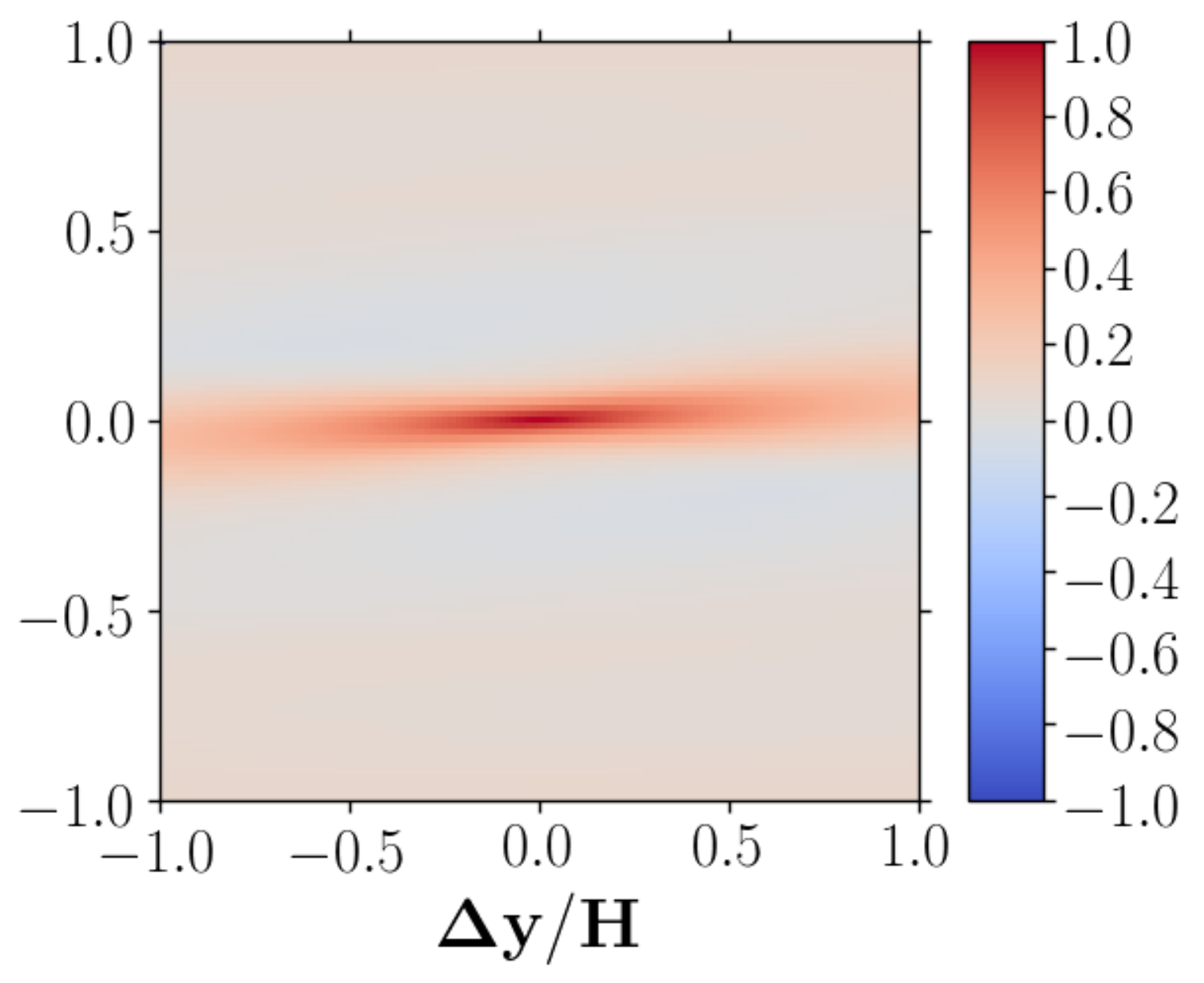}
\caption{Autocorrelation functions for the vertical velocity for the medium dead zone (top) and large dead zone (bottom).  Moving from left to right, the ACF is calculated using planes $|z|=$ 0 to 0.5, 1.0 to 1.5, 2.0 to 2.5 and 3.0 to 3.5.  Each plot is $2H$ on a side.  The large scale circulation modes can clearly be seen in the leftmost plots, while the plots on the right corresponding to the upper layers are consistent with the structure of MRI-driven turbulence in the ideal MHD limit, as seen
in other quantities and previous studies \cite[e.g.,][]{guan09}. }
\label{figACFvel}
\end{figure*}

To further quantify the turbulent structure as a function of height, we use autocorrelation functions (ACF) to 
characterize the degree of non-axisymmetry present in the turbulence. The 2D ACF in the $x$-$y$ plane for a quantity $f$ is defined as,

\begin{equation}
\text{ACF}(f(\Delta \vec{x}^{\,})) = \overline{ \bigg( \frac{\int f(t ,\vec{x}^{\,}) f(t ,\vec{x}^{\,} + \Delta \vec{x}^{\,}) d^2 \vec{x}^{\,}}{\int f(t ,\vec{x}^{\,})^2 d^2 \vec{x}^{\,}} \bigg) , }                     
\end{equation} 

\noindent where $\vec{x}^{\,}$ is the position in the $x$-$y$ plane, $\Delta \vec{x}^{\,}$ is the shift in this position, and the over bar indicates an average over time.  With this definition, the ACF is normalized by its maximum value at $\Delta \vec{x}^{\,}=0$.  The time averages are performed during the saturated state for each run (see table~\ref{tabResults}).  An ACF was calculated in the $x$-$y$ plane for each $z$ and time, and then averaged over the desired range of $z$ and time-interval.  Examples of time and volume averaged ACFs for the density perturbation and $v_z$ can be seen in Figures~\ref{figACFdensPert} and~\ref{figACFvel} respectively.

From the figures it is clear that the structure of the turbulence (or lack thereof) is a strong function of $z$, as expected for a disk with a dead zone. In the active zone the density perturbation de-correlates over a short distance scale, and the 
ACF is a small, significantly tilted ellipse, consistent with previous studies \citep{guan09,simon12}. Near the mid-plane the density fluctuations are almost axisymmetric, consistent with the dynamics in this region being dominated by 
spiral density waves. In an intermediate region near the lower parts of the active zone, the presence of the dead zone causes the turbulence not to look like ideal MRI even where $\Lambda \gg 1$.

\begin{figure}[h]
\centering
\includegraphics[width=\columnwidth]{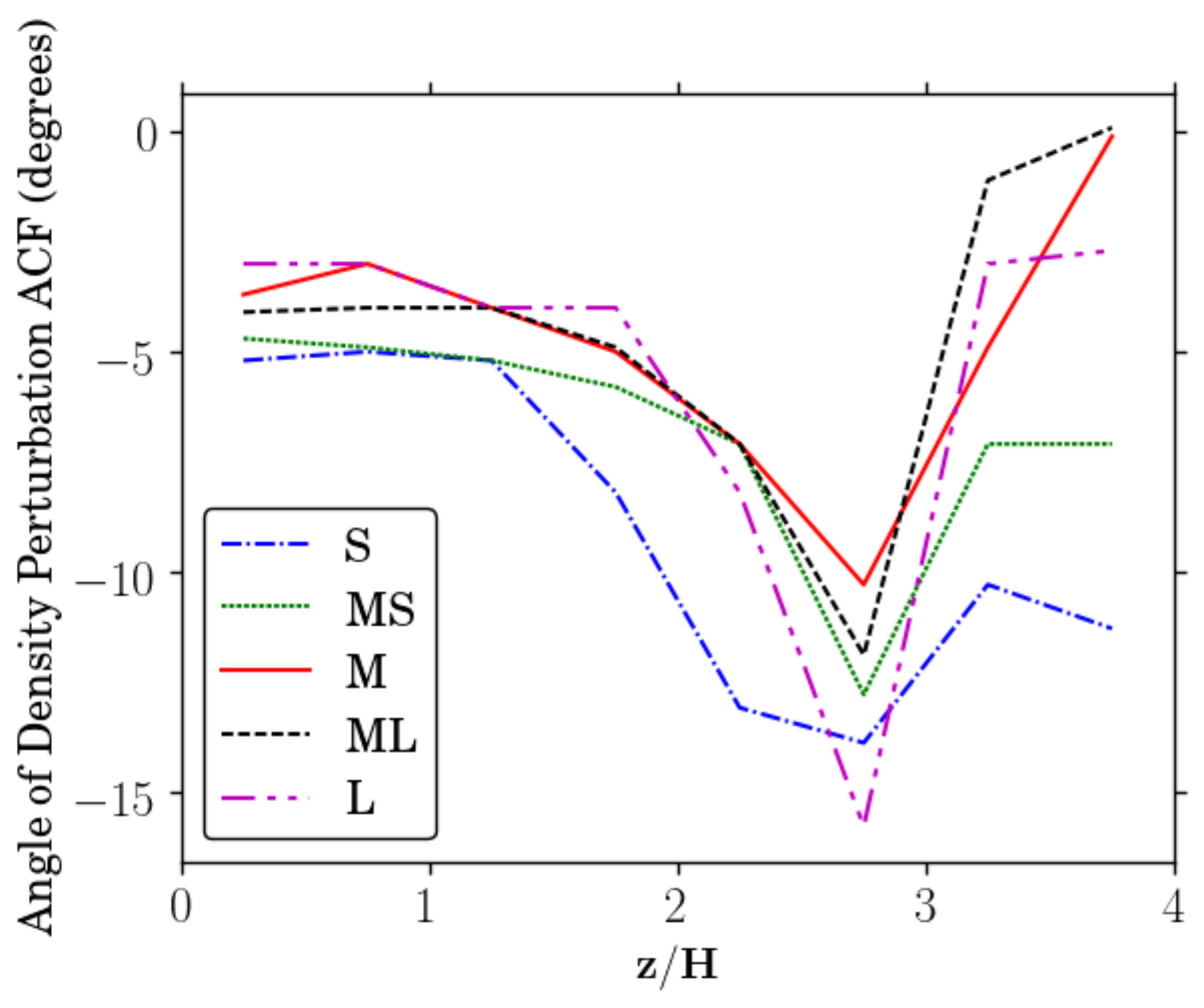}
\caption{The angle of the density perturbation ACF away from the y-axis for each of the dead zone runs as a function of $z$.}
\label{figAcfAngleProfile}
\end{figure}

The shape of the ACFs was quantified by fitting an ellipse to the density perturbation ACF.  The ellipse was fit such that it had the minimum possible area while containing at least 10 percent of the total value of the entire ACF.  
In the case of the ideal MRI, prior shearing box simulations show that the expected angle of the density perturbation ACF is about $-7^\circ$ offset from the y-axis \citep{guan09}.  This angle is plotted as a function of height for each run in Figure~\ref{figAcfAngleProfile}. In the active zone (higher than $z/H=2.0$ for all runs), the angle is near the expected value for ideal-MHD MRI turbulence, while in the dead zone the turbulent fluctuations are more closely aligned along the $y$-axis.  The degree of alignment generally increases as the dead zone thickness increases. We also note that the angle measured at the very outer part of the disc ($|z/H|>3.0$) is discrepant as compared to ideal-MHD expectations. This could be due to the influence of boundary conditions near the edge of the box, or because the magnetic pressure dominates the gas pressure in this region, quenching the MRI and thus changing the expected structure of turbulence in this region.

The ACFs of the magnetic field components are consistent with no magnetic activity in the dead zone and the expectations of the ideal MRI from \cite{guan09} in the active zone. However, the ACF of the vertical velocity displays distinctly different behavior.  In the active zone, the vertical velocity ACF has the modestly non-axisymmetric structure that is characteristic of turbulence in the active zone, as seen in the ACF of other quantities.  However, near the dead-zone mid-plane, there are coherent, large-scale structures in the vertical velocity. We analyze the properties of these structures further in the next section.

\subsection{Large scale velocity structures in the dead zone}
\label{sec_circulate}   

\begin{figure}[h]
\centering
\includegraphics[width=0.7\columnwidth]{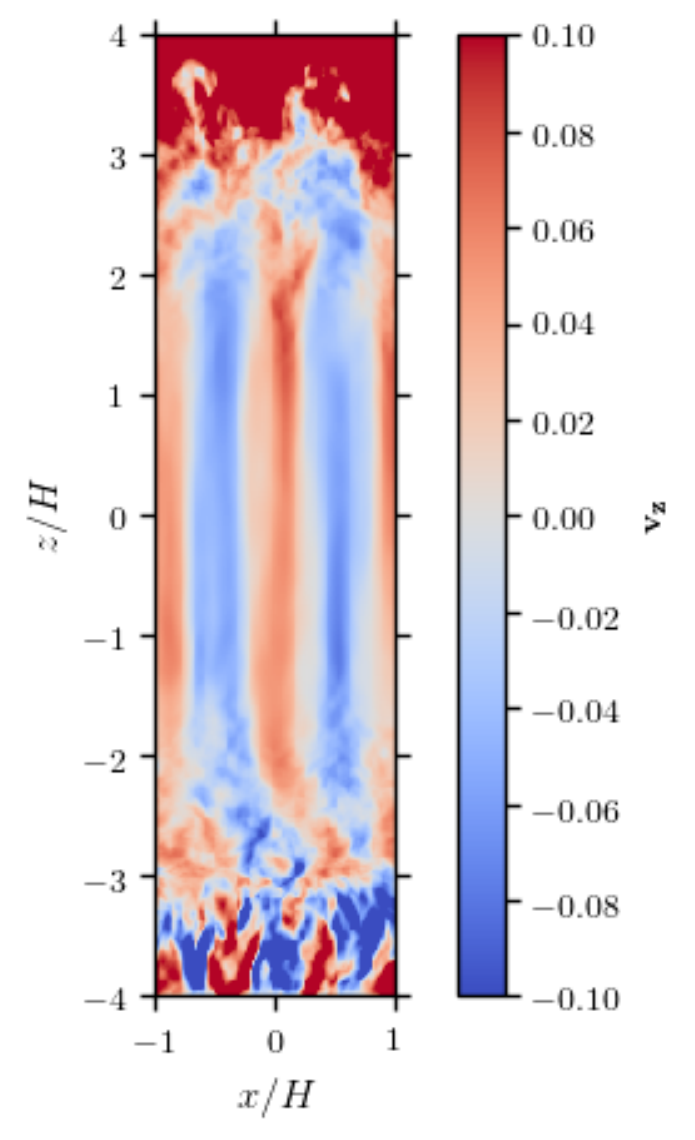}
\caption{A snapshot from the medium-sized dead zone run of the vertical velocity in the $x$-$z$ plane, averaged over $y$. The figure clearly depicts the presence of two 
vertically coherent updrafts / downdrafts, stretching throughout the dead zone region.}
\label{figVzLs}
\end{figure}

\begin{figure}[h]
\includegraphics[width=\columnwidth]{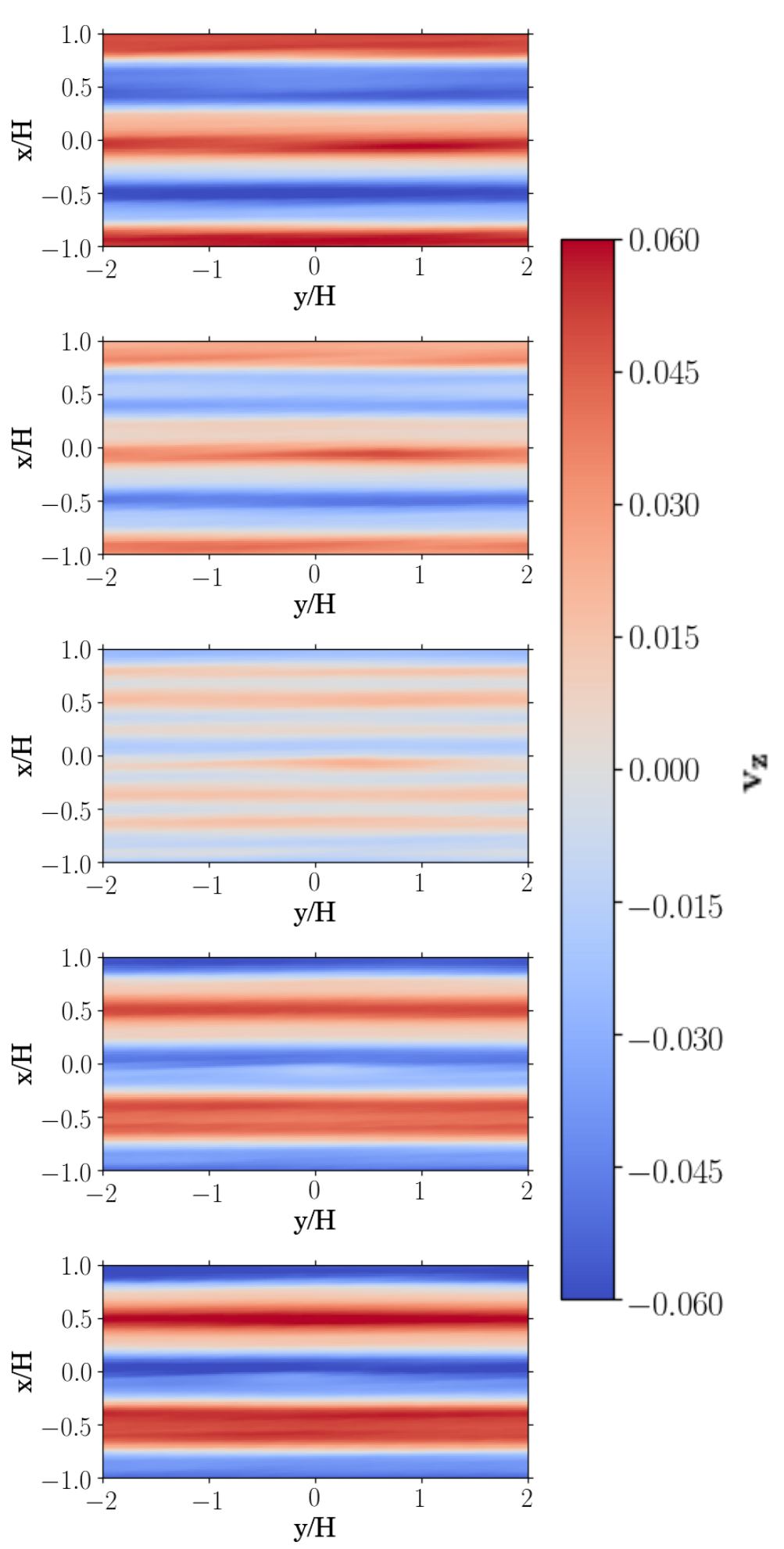}
\caption{The vertical velocity plotted for the $x$-$y$ plane at the mid-plane of the medium-sized dead zone run.  Time advances from top to bottom by 3.5 $\Omega^{-1}$ each panel, showing half of a cycle of this standing wave pattern.}
\label{figVzLs2}
\end{figure}

The structure of the large scale velocity field in the dead zone is shown in Figure~\ref{figVzLs} and 
Figure~\ref{figVzLs2}. Figure~\ref{figVzLs} shows a snapshot of the vertical velocity in the $x$-$z$ plane (averaged over $y$) for the medium sized dead zone simulation. Figure~\ref{figVzLs2} shows the vertical velocity in the $x$-$y$ plane at the mid-plane at a number of different time slices. In the example shown, the vertical velocity is dominated by a mode with a horizontal wavelength of $H$, which is coherent vertically between about $z = \pm2H$. Two of these modes fit within the radial extent of the shearing box. An inspection of the time sequence of snapshots at $z = 0$ shows that the large scale velocity structures take the form of a low frequency oscillatory meridional circulation. 
A Fourier analysis shows that the frequency is approximately $0.036 \Omega$, implying a period of about 4.5 orbits. This dominant (primary) mode is not the only such structure present. Figure~\ref{figVzLs2} shows half a cycle of the oscillation. When the dominant mode is at a minimum, a weaker (secondary) mode with wavelength $2H/7$ is apparent. We have examined how the kinetic energy is partitioned between small-scale and large-scale fluid motions as a function of height. As is clear visually, small scale kinetic energy dominates by a factor of a few in the active zone, while conversely large scale kinetic energy dominates by a factor of a few in the dead zone.

The spontaneous development of large scale velocity structure in the simulated dead zone is somewhat surprising, and we have made various checks to assess whether similar structures would be present in real disks. We first note that similar features were already seen in the shearing box simulations of \cite{oishi09}, which were run with a different code (Pencil). Those authors also found evidence for similar modes (though with reduced amplitude) in simulations where the isothermal equation of state was changed to ideal. We are therefore confident that the large scale circulation is not a numerical artifact of the {\sc Athena} code, though as we will discuss below, the manifestation of disk modes in the shearing box approximation can differ from those expected in a physical disk. We have also verified that the mid-plane of the simulated disk is stable to within a grid cell over the course of a vertical oscillation cycle. This implies that our use of a resistivity profile that is fixed in $z$, rather than being specified by column density directly, is unlikely to drive the modes unphysically.

We interpret the  standing wave pattern in Figure~\ref{figVzLs2} in terms of disk modes.
We ignore the effects of magnetic fields that play a key role in the active zone.
We apply the approximation that the dead zone occupies the full vertical extent of the
disk in computing the mode structure.
This approximation should approximately hold, since
the active zone involves only about 11\% of the disk mass and lies well outside
the  disk mid-plane region where the dead zone resides.
In the notation of \cite{lubow93} the implied wave angular frequency from Figure~\ref{figVzLs2} is
 $\omega \simeq 0.22 \Omega$. Since plots in Figure~\ref{figVzLs2}  are nearly 
independent of the azimuthal coordinate $y$, we take these modes to be axisymmetric.
The only axisymmetric modes that have  $\omega <  \Omega$ are r modes \citep{lubow93,ogilvie98}. 
These modes derive their support from inertial and rotation forces. For disks that behave adiabatically in the presence of vertical entropy gradients, buoyancy
forces also play a role in confining the mode near the disk mid-plane (see Figure~11 of \cite{lubow93}, who referred to these modes as g modes).
In the simulations of this paper, buoyancy plays no role, since the disk is strictly isothermal.

We suggest that the wave pattern in Figure~\ref{figVzLs2} is the lowest order (in terms of vertical structure) r mode.  This mode has
no vertical nodes in the vertical velocity. Its linear dispersion relation for an isothermal Keplerian 
disk follows from equation (54) of LP93
with $n=0$ (no vertical nodes in the vertical velocity), $\kappa=\Omega$ (which is equivalent to $\kappa_{\rm LP93} = 1$), and $\gamma=1$ (isothermal perturbations in an isothermal disk) and is given by
\begin{equation}
\omega = \frac{\sqrt{2}\, \Omega}{2} \sqrt{ 2 + K^2 - |K| \sqrt{4 + K^2} },
\label{omega}
\end{equation}
where $K = k_{\rm x} c_{\rm s}/\Omega= k_{\rm x} H/\sqrt{2}$ is the scaled radial wavenumber.

The perturbations caused by the linear wave depend on a scale factor that
we take to be the vertical velocity amplitude $W$. This amplitude depends on the strength
of forcing from the active layer.
The standing wave properties are given by
\begin{eqnarray}
v_{\rm x} (x, z, t) &=& \sqrt{2} W   \left(\frac{\omega z}{ \Omega H}  \right) \sin{( k_{\rm x} x)} \cos{(\omega t)},
\label{eq:vx}\\
v_{\rm y} (x, z, t) &=& -\frac{\sqrt{2}}{2}  W  \frac{ z}{H}  \sin{(k_{\rm x} x)} \sin{(\omega t)}, \\
v_{\rm z} (x, z, t) &=& W  \cos{(k_{\rm x} x)} \cos{(\omega t)} ,\\
p(x, z, t) &=&  2 \,  \left( \frac{W\, \omega\, z}{H^2 \Omega^2} \right)\, p_0 \exp{(-z^2/H^2)}  \nonumber \\
& \times &  \cos{(k_{\rm x} x)} \sin{(\omega t)},
\label{eq:p}
\end{eqnarray}
where  $p_0$ is the unperturbed disk pressure at mid-plane.
Note that we have selected a phasing in which $v_{\rm z}(0,z,0)=W$.
In the low frequency limit for fixed $W$, the pressure perturbation vanishes, while the wavenumber diverges in lowest order as $K = \Omega/\omega$. The velocity divergence does not vanish in this limit and is finite.

The radial boundary conditions used in the simulations 
imply that there is a discrete set of radial wave numbers in the disk with
\begin{equation}
k_{\rm x} =  \frac{\sqrt{2} K}{H}=  \frac{2 \pi \ell}{L_{\rm x}}
\end{equation}
for integer $\ell$ and $L_{\rm x} = 2H$ in the simulations presented here.
The discreteness is due to the limited box size.
The wave structures given by Equations (\ref{eq:vx}) - (\ref{eq:p}) satisfy this boundary condition for the radial wavenumbers found in the simulations.  The  primary mode with radial wavelength $H$ and the secondary mode with radial wavelength $2 H/7$ identified in Figure~\ref{figVzLs2}
correspond to $\ell=2$ and 7 respectively.  The level of excitation of these particular modes is determined by the overlap of their space and time dependences with these dependences for the turbulence.

For the primary mode that has $\ell=2$, wavelength $H$, and the wavenumber $K=\sqrt{2}\, \pi$, the r mode dispersion relation
given by Equation (\ref{omega}) evaluates to $\omega = 0.2147 \Omega$, 
in good agreement with the angular frequency $\omega \simeq 0.22 \Omega$
that is determined by the time changes in Figure~\ref{figVzLs2} for the same wavelength $H$ disturbance. 
The agreement gives us confidence that we have properly identified this dead zone disturbance as being due to the lowest order r mode.  

We expect that the other (secondary) wave mode with radial wavelength $2H/7$ to be roughly similar to the primary r mode that has radial wavelength $H$ and no vertical nodes in the vertical velocity, $n=0$, 
since both modes are excited by the active layer turbulence.
The $n=1$ r mode has zero vertical velocity at the disk midplane at all times
and therefore cannot explain the presence of a vertical velocity in Figure~\ref{figVzLs2}. (The $n=1$ mode may exist, but is
is not detectable in Figure~\ref{figVzLs2}.)
A possible candidate for this secondary mode seen in Figure~\ref{figVzLs2} is the $n=2$ r mode which has a nonzero vertical velocity at the mid-plane.  The dispersion relation of Equation (\ref{omega}) applied to this mode predicts a lower angular frequency that is about half the frequency of the $n=0$ mode.  Due to the substantial frequency difference between the primary and secondary modes, we expect a substantial phase difference between the modes when the primary mode has zero vertical velocity.  We then expect the secondary mode to have a nonzero vertical velocity at this time, as seen in the middle panel of Figure~\ref{figVzLs2}. 

We speculate on how the shearing box simulations might carry over to a full disk.
For a full disk, the radial wave numbers would be continuous rather than  discrete.
Furthermore, the waves would not generally be standing waves. A mix of inwardly
and outwardly propagating r mode waves would be excited by turbulence in the active layer. 
An outwardly propagating  r mode with $n=0$ becomes an outwardly propagating p (pressure) mode when it propagates past  its corotation radius that occurs
where $\Omega(r) = \omega$ \citep{lubow93}.  An inwardly propagating wave  remains as an r mode.
 An outwardly propagating  r mode with $n>0$ reflects at corotation 
and becomes an inwardly propagating r mode.  These waves would be subject
to damping by shocks and interactions with turbulence.

The flow pattern of an r mode is
quite different for a disk with buoyancy. Buoyancy acts to
vertically confine r modes near the disk mid-plane. The confinement condition on $z$ is that $N(z) < \omega$, where $N$ is the vertical
buoyancy frequency.
For $\gamma = 1.4$, the r mode extends over height $|z| <
1.87 \omega c_{\rm s}/\Omega^2$, which for $\omega=0.2 \Omega$, implies $|z| < 0.3 H$. The extent of driving the mid-plane motions could also be affected, if the base of the active region lies above this height. For any
reasonably sized dead zone, we would then expect a major change in the vertical flow properties of a disk that undergoes adiabatic perturbations.

\section{Implications for Particle Dynamics}
Proceeding under the assumption that the wave modes seen in the simulations are present in disks, we now ask what impact they would have on the distribution of solid particles. As \cite{oishi09} noted, the low frequency and non-turbulent nature of the vertical velocity field means that there is no direct effect on particle settling for the small 
(dimensionless stopping time $\tau \ll 1$) particles generally assumed as progenitors for planetesimal formation. However, a low frequency vertical flow could have the effect of ``buckling" a vertically settled particle layer, such that the mid-plane of the particle layer coincides with the mid-plane of the gas disk only for a fraction of an oscillation cycle. For realistic amplitudes, it is not clear that buckling of a mid-plane particle layer would have directly observable effects. A buckled particle layer would, however, suppress the rate of pebble accretion \citep{lambrechts12} on to larger bodies that are aerodynamically or gravitational damped to the mid-plane of the gas disk.

To estimate the potential importance of this effect, we first determine the expected amplitude of the vertical oscillations that would be experienced by a settled particle layer. 
To first order, the circulation modes discussed in \S\ref{sec_circulate} have a sinusoidal time dependence at a given point in the mid-plane,

\begin{equation}
v_z=v_{z,0}\cos(\Omega_c t)                
\end{equation}

\noindent where $\Omega_c$ is the oscillation frequency of the circulation mode and $v_{z,0}$ is the peak vertical velocity of the mode at the mid-plane.  Under this assumption, the maximum displacement of a particle that is perfectly coupled to the gas is given by

\begin{equation}
\Delta z_c = \frac{v_{z,0}}{\Omega_c}.                
\end{equation}

\noindent The circulation mode parameters and this estimate of the displacement are given in Table~\ref{tabCorrugation} for each different dead zone size. Because of the low frequencies involved, the modest velocities (of the order of 
$0.1 c_s$) are able to induce vertical displacements that are of the order of $H$ for all of the dead zone sizes considered.

The buckling of the particle layer will affect the efficiency of pebble accretion if the amplitude exceeds the settled thickness of the layer. For particles with dimensionless stopping time $\tau_s$ the thickness of this particle layer can be written as,   

\begin{equation}
\frac{H_d}{H} = \sqrt{\frac{D_{d,z}}{\tau_s}}     
\label{eqnParticleH}  
\end{equation}

\noindent where $H_d$ is the scale height of the layer and $D_{d,z}$ is the dimensionless vertical dust diffusion coefficient.  For particles that are well coupled to the gas ($t_s \ll \Omega^{-1}$), this diffusion coefficient is the same as the vertical gas diffusion coefficient $D_{g,z}$, given by,  

\begin{equation}
D_{d,z} = D_{g,z} = \frac{\Omega}{c_s^2} \int_{0}^{\infty} \langle v_z(t)v_z(0) \rangle {\rm d}t.          
\label{eq_D}  
\end{equation}

\noindent The integrand is the auto-correlation function of the vertical gas velocity \citep{zhu15}.  In order to calculate this parameter, a high frequency (10 snapshots per $\Omega^{-1}$) 3D sample of data was output from the medium-sized dead zone run over 5 orbits.  To calculate the diffusion coefficient without contamination from the circulation modes themselves, the circulation mode must be subtracted from the velocity.  Given that the mode is roughly constant across $y$ and $z$ near the mid-plane, the circulation velocity $v_{z,c}$ was defined to be,

\begin{equation}
v_{z,c}(x,t) = \frac{\langle \rho(x,y,z,t)  v_z(x,y,z,t) \rangle_{y,z}}{\rho_{\rm avg}},
\end{equation}

\noindent where $\rho_{avg}$ is the average density over the mid-plane region being considered.  The random velocity $v_{z,\text{random}}$ entering into equation~(\ref{eq_D}) is then taken to be,

\begin{equation}
v_{z,{\rm random}} = v_{z}-v_{z,c}.
\end{equation}

\begin{figure}[h]
\centering
\includegraphics[width=\columnwidth]{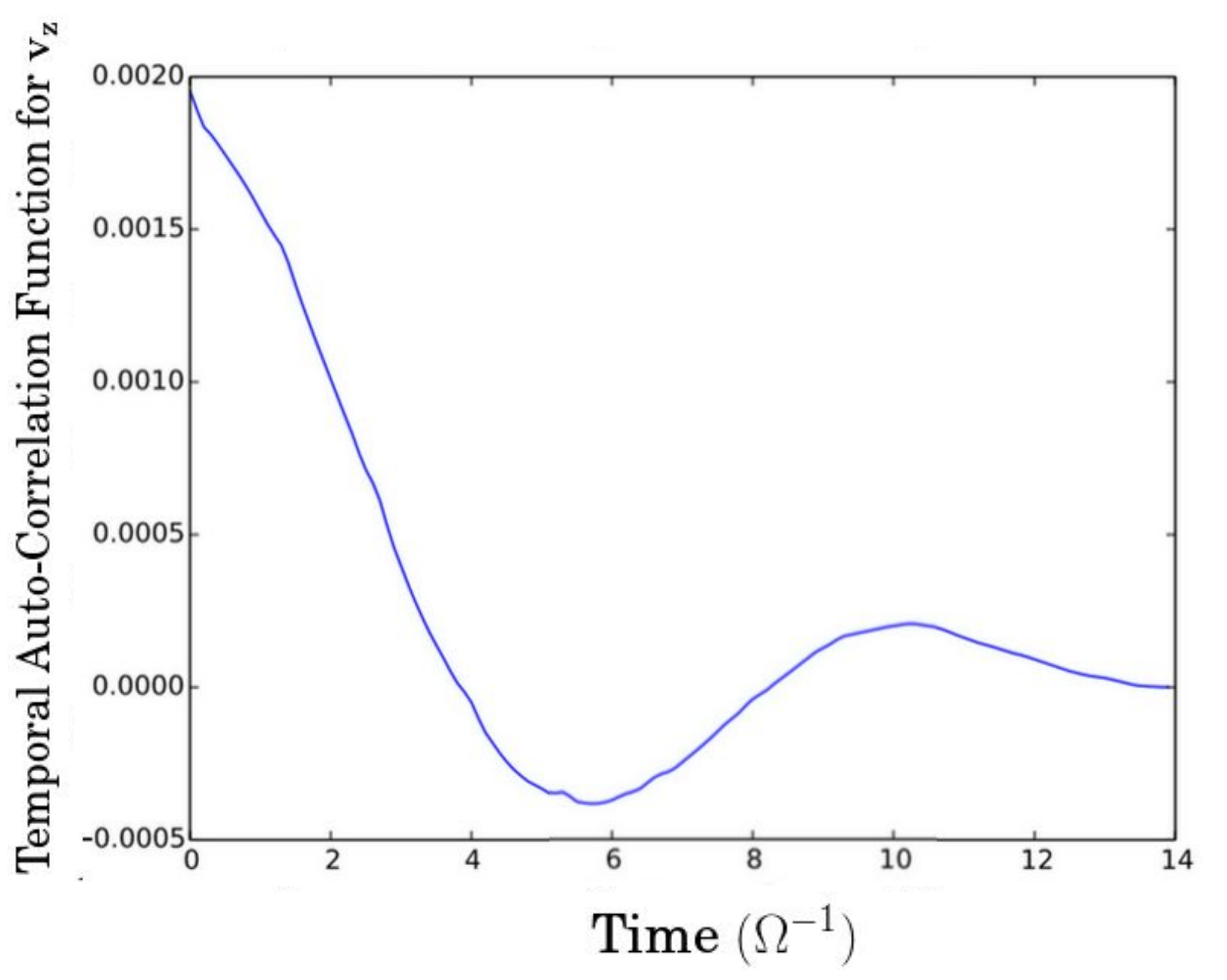}
\caption{The average auto-correlation function of the random vertical velocity for all points $-0.5<z/H<0.5$ as a function of the lag time.  The vertical diffusion parameter is the area under this curve.  }
\label{plotCCF}
\end{figure}

Using this random vertical velocity, the average auto-correlation function over all points near the mid-plane ($-0.5<z/H<0.5$) was calculated. This function, shown in Figure~\ref{plotCCF}, is well-behaved, and when integrated yields a dimensionless vertical diffusion parameter of $7.2 \times 10^{-3}$ for the middle sized dead zone. We roughly estimate this parameter for the other runs by assuming that $D_{d,z}$ scales with $\delta v^2$ at the mid-plane. Using equation~\ref{eqnParticleH}, the scale height of the particle layer can now be determined for a given stopping time $\tau_s$. 

If the displacement due to the circulation mode is comparable to the scale height of the particle layer, the mode will have a significant effect on the structure of the layer.  This criterion can be used to define a critical stopping time $\tau_{s,crit}$ at which $H_d$ and $\Delta z_c$ are equal,

\begin{equation}
\tau_{s,crit} = \frac{D_{d,z}}{(\Delta z_c/H)^2}   
\end{equation} 

\noindent These critical stopping time values are shown in Table~\ref{tabCorrugation}.  The results indicate that, in general, particle layers with $\tau_s \geq 10^{-3}$ will be significantly perturbed by these circulation modes.

\begin{table}
\centering
 \begin{tabular}{|l | c | c | c | c | c |} 
 \hline
 DZ Size & $v_{z,0}$ & $\Omega_c$ & $\Delta z_c/H$ & $D_{d,z}$ & $\tau_{s,\text{crit}}$ \\  
 \hline
 S & 0.10 & 0.039 & 2.56 & $1.4\times10^{-2}$ & $2.2\times10^{-3}$ \\
 \hline
  MS & 0.08 & 0.038 & 2.11 & $1.1\times10^{-2}$ & $2.6\times10^{-3}$ \\
 \hline
  M & 0.06 & 0.036 & 1.67 & $7.2\times10^{-3}$ & $2.6\times10^{-3}$ \\
 \hline
 ML & 0.03 & 0.024 & 1.25 & $2.8\times10^{-3}$ & $1.8\times10^{-3}$ \\
 \hline
 L & 0.014 & 0.024 & 0.58 & $4.6\times10^{-4}$ & $1.4\times10^{-3}$ \\
 \hline
\end{tabular}
\caption{Parameters of the circulation modes, vertical particle diffusion parameters, and critical $\tau_s$ values for each run.  }
\label{tabCorrugation}
\end{table}

\section{Conclusions}
We have used local numerical simulations to study the structure of Ohmic dead zones, that may be relevant to 
accretion in protoplanetary disks \citep{gammie96} and in the outer regions of dwarf novae \citep{gammie98}. 
By setting up a simplified, yet well-defined model --- that ignores the role of other non-ideal processes and has a sharp 
transition between magnetically active and inactive zones --- we were able to run long duration simulations at high 
resolution in order to precisely quantify how dead zone properties scale with the thickness of the dead zone. We found that:
\begin{itemize}
\item[(i)]
The Reynolds stress, which dominates transport near the mid-plane, scales strongly with the thickness of the 
dead zone. For modest dead zones, in which the ratio of magnetically active to inactive column is of the order of 
0.1 or higher, we find $\alpha \sim 10^{-4}$. Thicker dead zones, in which the active column is only of the order of 
1\% of the total mass, have negligible mid-plane stresses. As a result, small dead zones can have accretion rates that are a significant fraction of the active zone accretion rate ($10$ to $30\%$), while larger dead zones have accretion rates around $1\%$ of the active zone rate.  This scaling supports models in which mass can 
accumulate in the dead zone region, building up a reservoir of material that can potentially accrete 
rapidly \citep{armitage01,zhu09,martin11}.
\item[(ii)]
The structure of turbulence in the MRI-active region, at high $z$, resembles that seen in ideal MHD 
simulations of fully active disks \citep{guan09,simon12}. The hydrodynamic component of the turbulence  
becomes more axisymmetric and less efficient at angular momentum transport towards the mid-plane, 
especially for thick dead zones.
\item[(iii)]
The time scale for saturation of the kinetic energy in the dead zone is long (reaching $\sim 500$ orbits 
for the thickest dead zone studied). Most of the kinetic energy is contained in large-scale, non-turbulent 
fluid motions, which take the form of an oscillatory meridional circulation that is extended in the 
vertical direction. Similar fluid motions were observed in independent simulations by 
\cite{oishi09}. We identify this motion with the lowest order r mode expected in a purely hydrodynamic 
disk model \citep{lubow93}.
\item[(iv)]
We expect the properties of the large-scale fluid motions to depend upon the nature of the perturbations 
(isothermal or adiabatic), and upon where vertical driving of the mode occurs. Moreover, the standing waves 
observed in our shearing box simulations would likely be replaced by inwardly and outwardly propagating waves 
in a global disk model. However, 
if models similar to those we observe are excited to significant amplitudes in more physical models of protoplanetary disks, the 
resulting buckling of settled particle layers could reduce the efficiency of pebble accretion for small ($\tau \ll 1$) particles. 
\end{itemize}

Although purely Ohmic dead zones are an idealization, the inclusion of 
non-ideal physics in disk simulations frequently leads to vertical stratification of turbulent properties, and regions of the 
disk where mid-plane MHD stresses are negligible, as we have seen here.  Further simulations will be needed 
to determine if our results for the scaling of Reynolds stresses, and for the excitation of large scale 
fluid flows, carry over under more realistic conditions. \\

\acknowledgments
We acknowledge support from NASA through grants NNX13AI58G, NNX14AB42G and NNX16AB42G (P.J.A), from the NSF through grant AST 1313021 (P.J.A.), and from grant HST-AR-12814 (P.J.A.) awarded by the Space Telescope Science Institute, which is operated by the Association of Universities for Research in Astronomy, Inc., for NASA, under contract NAS 5-26555. J.B.S.'s support was provided in part under contract with the California Institute of Technology (Caltech) and the Jet Propulsion Laboratory (JPL) funded by NASA through the Sagan Fellowship Program executed by the NASA Exoplanet Science Institute. 
This work utilized the Janus supercomputer, which is supported by the National Science Foundation (award number CNS-0821794), the University of Colorado Boulder, the University of Colorado Denver, and the National Center for Atmospheric Research. The Janus supercomputer is operated by the University of Colorado Boulder. S.H.L. acknowledges support from NASA grant NNX11AK61G.

\end{document}